\documentclass[11pt,a4paper]{article}
\usepackage[T1]{fontenc}
\usepackage[utf8]{inputenc}
\usepackage{lmodern}
\usepackage{amsmath,amssymb,amsfonts}
\usepackage{graphicx}
\usepackage{booktabs,tabularx,array}
\usepackage{microtype}
\usepackage{url}
\usepackage[hidelinks]{hyperref}
\usepackage{geometry}
\usepackage{float}
\usepackage{enumitem}
\usepackage{setspace}
\usepackage{upgreek}
\newcolumntype{C}{>{\centering\arraybackslash}X}
\begin{document}
\title{Coupled-Channel Spectral Theory for a Non-Separable Geometry: Normal Modes and Two-Boundary Response in the Rotating AdS-Teo Wormhole}
\author{Ramesh Radhakrishnan\thanks{Corresponding author: \texttt{ramesh\_radhakrishna1@baylor.edu}}\textsuperscript{1,2}
\and William Julius\textsuperscript{1,2,3}
\and Gerald Cleaver\textsuperscript{1,2}}
\date{}
\maketitle
\begin{center}
\small
\textsuperscript{1}Department of Physics, Baylor University, Waco, TX 76798, USA\\
\textsuperscript{2}Early Universe, Cosmology and Strings (EUCOS) Group, Center for Astrophysics, Space Physics and Engineering Research (CASPER), Baylor University, Waco, TX 76798, USA\\
\textsuperscript{3}College of Science and Engineering (COSE), St. Cloud State University, St. Cloud, MN 56301, USA
\end{center}
\begin{abstract}

We investigate scalar perturbations of a rotating asymptotically
anti-de Sitter (AdS)-Teo traversable wormhole with a controlled
non-separable angular deformation.  The geometry retains a regular
wormhole throat and the required AdS asymptotics, while an explicit
quadrupolar deformation generates angular-channel coupling in the
intermediate region.  A sufficient condition is derived for an
ergoregion-free parameter regime in which the spectral analysis is
performed.  Projecting the geometry-derived Klein--Gordon equation
onto spherical harmonics yields a matrix-valued Sturm--Liouville
system and an associated quadratic operator pencil.  Throat
regularity together with normalizable AdS boundary conditions leads
to a determinant quantization condition for the discrete normal-mode
spectrum.  Two-, four-, and six-channel calculations demonstrate
systematic numerical convergence of the retained low-lying
normal-mode frequencies under enlargement of the angular basis.
The complete finite generalized eigenspectrum is also examined
without imposing a near-reality selection criterion, and no growing
scalar mode is found within the ergoregion-free parameter range and
numerical resolutions studied.  The six-channel rotation
continuation exhibits a finite interior level-repulsion feature
accompanied by collective redistribution of the multichannel
eigenvectors.  The same coupled spectral framework formally defines
a matrix-valued two-boundary response whose poles are selected by the
normal-mode matching condition; the response plots presented here
illustrate this structure using the reduced two-channel effective
model rather than a numerical reconstruction of the full six-channel
response.  We further derive the geometry-dependent short-distance
Hadamard subtraction and identify the finite inter-channel structure
entering a truncated local quantum mode sum, without claiming a
complete numerical evaluation of
$\langle\Phi^2\rangle_{\rm ren}$.  Together, these results establish a
geometry-to-spectrum framework in which the spacetime geometry
determines the coupled operator, the global boundary conditions
determine its normal-mode spectrum, and the same coupled spectral
framework organizes the associated asymptotic response and local
quantum mode-sum structure.
\end{abstract}

\section{Introduction}
\label{sec:introduction}

Quantum field theory in curved spacetime provides the natural framework
for studying how geometry, symmetry, topology, and~boundary conditions
determine the propagation and spectral structure of classical and
quantum fields in gravitational \mbox{backgrounds~\cite{Parker1969,Fulling1973,BirrellDavies,WaldQFTCS}}.
In asymptotically anti-de Sitter (AdS) spacetimes, the~timelike
conformal boundary converts wave propagation into a global
boundary-value problem and, under~reflective boundary conditions,
leads naturally to a discrete normal-mode \mbox{spectrum
~\cite{BreitenlohnerFreedman1982,IshibashiWald2004}}.
For geometries with more than one asymptotic region, the~same global
spectral problem also carries information about propagation and
correlations between distinct boundaries
~\cite{Maldacena1998,Witten1998,SonStarinets2002}.

Traversable wormholes provide a useful setting in which these effects can
be studied explicitly.  Beginning with the Morris--Thorne construction
and its subsequent \mbox{extensions
~\cite{MorrisThorne1988,Visser1995}}, wormhole geometries have been used to
investigate global causal structure, wave propagation, semiclassical
effects, and~the role of nontrivial topology.  Rotating traversable
wormholes introduce additional structure through frame dragging, and~the
Teo metric provides a standard stationary and axisymmetric model of such
a geometry~\cite{Teo1998}. Recent work on Ellis--Bronnikov wormholes provides an important point of comparison for this construction.  Blazquez-Salcedo~et~al.\ constructed explicit traversable Ellis wormholes with anti-de Sitter asymptotics supported by a phantom scalar field, demonstrating that two-ended AdS wormhole geometries can arise as solutions of a specified gravitational matter system~\cite{BlazquezSalcedo2021AdS}.  Our construction has a different purpose: we retain a rotating Teo-type throat and prescribe a phenomenological asymptotically AdS completion, with~the effective stress-energy tensor defined by the resulting geometry rather than derived from a microscopic matter~Lagrangian.

Rotating Ellis--Bronnikov wormholes have also been studied through increasingly
detailed perturbative analyses.  Azad~et~al.\ investigated radial perturbations
in the slow-rotation expansion~\cite{Azad2024SlowRotation,Azad2024SecondOrder}, while Khoo~et~al.\ extended
the analysis to rapidly rotating configurations and their quasinormal spectra
~\cite{Khoo2024QNM}.  Subsequent work examined quadrupolar perturbations of
slowly spinning Ellis--Bronnikov wormholes~\cite{Azad2025Quadrupole}.  These studies concern perturbations and stability of dynamically supported wormhole solutions.  The~problem considered here is complementary: the background geometry is prescribed, the~scalar field is a probe, and~the principal objective is to derive the non-separable geometry-induced angular coupling as a global matrix spectral~problem.

A central simplification in many rotating systems is complete
separability.  In~Kerr spacetime, hidden symmetry associated with the
Carter constant and the corresponding rank-two Killing tensor permits
the field equations to be decomposed into independent radial and angular
problems~\cite{Carter1968,Teukolsky1973,Chandrasekhar1983}.
Stationarity and axisymmetry alone, however, do not guarantee such a
structure.  Once the metric possesses genuine angular dependence, the~scalar equation may instead organize into interacting angular channels,
and the appropriate dynamical object becomes a matrix-valued radial
operator.

The purpose of this work is to construct and analyze this coupled
spectral problem explicitly for a rotating asymptotically AdS-Teo
wormhole.  We do not derive the wormhole from a specified microscopic
matter Lagrangian.  The~geometry is treated as a phenomenological fixed
background, with~the effective stress-energy tensor defined implicitly
through the Einstein equations with negative cosmological constant.  The~field considered throughout is a probe~scalar; the analysis,
therefore, concerns scalar-wave propagation on the prescribed background,
not gravitational perturbations or the dynamical stability of the
wormhole~itself.

The construction is organized by symmetry.  Stationarity and axisymmetry
preserve the Killing symmetries associated with time translations and
azimuthal rotations, while reflection symmetry across the throat permits
a parity decomposition of the global problem.  Complete angular
separability is then broken in a controlled manner through an explicit
$\theta$-dependent deformation of the metric.  The~deformation is chosen
to vanish at the throat and asymptotically, thereby preserving the
required global structure while generating genuine off-diagonal
couplings between spherical-harmonic channels.  Non-separability is,
therefore, generated by the geometry itself rather than imposed through
an effective matrix~model.

The analysis addresses four questions.  First, can one construct a
rotating asymptotically AdS-Teo background that interpolates consistently
between a regular wormhole throat and two AdS boundaries, admits a
controlled non-separable deformation, and~possesses an ergoregion-free
parameter regime?  Second, once complete separability is lost, what is
the correct global operator formulation of the scalar problem and how
are its discrete normal modes quantized?  Third, does the finite
coupled-channel approximation converge as additional angular channels
are retained, and~which spectral signatures survive this convergence?
Finally, how does the same global spectral framework organize the
formal two-boundary response and the coupled-channel structure entering
local \mbox{quantum observables?}

For the first question, we construct an explicit asymptotically AdS
radial sector together with a localized quadrupolar deformation of the
lapse.  The~resulting geometry retains the regular Teo throat, approaches
the required AdS asymptotics at both ends, and~provides an explicit
geometrical source of angular-channel coupling.  An~analysis of
$g_{tt}$ yields a sufficient condition for the absence of an ergoregion,
and all subsequent spectral calculations are restricted to this
admissible~regime.

For the second question, the~Klein--Gordon equation is derived directly
from the deformed geometry and expanded in spherical harmonics without
assuming radial-angular separability.  Angular projection produces a
matrix-valued Sturm--Liouville differential system whose coefficient
matrices are determined by the spacetime geometry.  The~corresponding
Hilbert space and associated weighted bilinear form, operator domain, AdS boundary conditions, and~frequency-dependent quadratic operator pencil are specified explicitly. Regularity through the throat together with the two asymptotic AdS boundary conditions then produces a geometry-derived matching matrix and the determinant quantization condition for the discrete normal-mode
spectrum. The~self-adjoint differential operator and the associated
quadratic spectral pencil are kept conceptually distinct throughout the
analysis.

For the third question, the~same geometry-derived coupled problem is solved in
successively enlarged angular spaces.  Two-, four-, and~six-channel calculations
show systematic convergence of the retained low-lying normal-mode frequencies
as additional channels are included. The~finite-dimensional systems, therefore, provide systematic Galerkin
approximations to the underlying infinite coupled-channel problem rather than independent phenomenological models. Within~the resulting six-channel realization, continuation in the rotation parameter reveals a finite interior level-repulsion feature accompanied by collective redistribution of the multichannel~eigenvectors.

The fourth question follows from the same global operator.  The~asymptotic solutions define formally a matrix-valued two-boundary
response whose poles are selected by the same matching condition that
determines the bulk normal-mode frequencies.  The~diagonal boundary
blocks describe same-boundary response, while the off-diagonal blocks
encode propagation between the two asymptotic AdS regions and retain
the coupled angular-channel structure
~\cite{SonStarinets2002,KaminskiOperatorMixing2010}.  In~this
work, the~full response matrix is not numerically reconstructed from
the six-channel bulk solutions; the response plots instead illustrate
the corresponding pole structure within the reduced two-channel
effective model.  The~same coupled-mode framework also enters the
formal Hadamard construction of local quantum observables.  Using
geometry-dependent temporal point splitting
~\cite{Christensen1976,DecaniniFolacci2008}, we identify the local
short-distance subtraction and the finite inter-channel structure of
the truncated mode representation, without~claiming a complete
numerical evaluation of $\langle\Phi^2\rangle_{\rm ren}$.

The near-throat operator structure is also examined as a local
characterization of the dynamics.  It provides useful information about
the symmetry organization of the throat region but is not the principal
result of this work.  The~main construction is global: the geometry
determines the coupled differential operator; the operator and AdS
boundary conditions determine the determinant spectrum developed in
Section~\ref{sec:global_spectral}; and the same coupled spectral
framework organizes the formal two-boundary response and the channel
structure entering local quantum mode sums developed in
Section~\ref{sec:boundary_quantum}.

Within this framework, the~normal modes are collective excitations of
interacting angular channels rather than independent single-harmonic
oscillations.  The~same geometry-induced coupling appears directly in
the six-channel spectral flow and organizes the reduced two-boundary
response and the finite-channel interference structure entering the
local quantum mode sum.  The~rotating AdS-Teo wormhole, therefore,
provides a concrete setting in which geometry, controlled symmetry
breaking, global boundary conditions, and~operator structure can be
followed continuously from the spacetime metric to the coupled
normal-mode spectrum and its associated response~framework.

The remainder of this paper is organized as follows.
Section~\ref{sec:geometry_scalar} constructs the rotating asymptotically
AdS-Teo geometry, develops the proper-distance formulation, derives the
coupled scalar-field equations, and~examines the local near-throat
operator structure. Section~\ref{sec:global_spectral} formulates the
global coupled spectral problem, specifies the operator and boundary
conditions, and~derives the determinant quantization condition for the
normal-mode spectrum. Section~\ref{sec:numerical_spectrum} presents the
numerical implementation, convergence analyses, complete complex-spectrum
audit, and~rotation-dependent multichannel spectral evolution.
Section~\ref{sec:boundary_quantum} develops the formal two-boundary
response framework and examines the associated quantum observables.
Finally, Section~\ref{sec:discussion_conclusions} summarizes the main
results, limitations, and~broader implications of the coupled-channel
spectral~framework.

\section{Rotating AdS-Teo Geometry and Coupled Scalar~Dynamics}
\label{sec:geometry_scalar}

We consider a stationary, axisymmetric rotating wormhole of the form
\begin{equation}
ds^2
=
-N^2(r,\theta)\,dt^2
+
\frac{dr^2}
{\left(1-\dfrac{b(r)}{r}\right)\mathcal{A}(r)}
+
r^2K^2(r,\theta)
\left[
d\theta^2
+
\sin^2\theta
\left(d\phi-\Omega(r,\theta)\,dt\right)^2
\right].
\label{eq:ads_teo_metric}
\end{equation}

The wormhole throat is located at $r=r_0$, where
\begin{equation}
b(r_0)=r_0,
\label{eq:throat_condition}
\end{equation}
and traversability requires the flare-out condition~\cite{MorrisThorne1988,Visser1995,Teo1998}
\begin{equation}
b'(r_0)<1.
\label{eq:flareout_condition}
\end{equation}

The spacetime is treated throughout as a phenomenological fixed
background rather than as a solution derived from a specified matter
Lagrangian.  The~effective source required to support the geometry is,
therefore, defined by
\begin{equation}
T_{\mu\nu}^{\rm eff}
=
\frac{1}{8\pi}
\left(
G_{\mu\nu}
+
\Lambda g_{\mu\nu}
\right),
\qquad
\Lambda=-\frac{3}{L^2},
\label{eq:effective_stress_tensor}
\end{equation}
where $L$ is the AdS curvature radius.  No microscopic interpretation of
$T_{\mu\nu}^{\rm eff}$ is~assumed.

For the background considered here we take
\begin{equation}
N(r,\theta)
=
N_0(r)+\mathcal{D}(r,\theta),
\qquad
N_0(r)
=
\sqrt{\mathcal{A}(r)}
=
\sqrt{
1+\frac{r^2-r_0^2}{L^2}
},
\label{eq:lapse_definition}
\end{equation}
together with
\begin{equation}
K(r,\theta)=1,
\qquad
\Omega(r,\theta)=\Omega(r)=\frac{2a}{r^3},
\qquad
b(r)=\frac{r_0^2}{r},
\label{eq:metric_functions}
\end{equation}
and
\begin{equation}
\mathcal{A}(r)
=
1+\frac{r^2-r_0^2}{L^2}.
\label{eq:A_definition}
\end{equation}

At the throat,
\begin{equation}
\mathcal{A}(r_0)=1,
\qquad
N_0(r_0)=1,
\label{eq:throat_A_N}
\end{equation}
while the shape function gives
\begin{equation}
b'(r)
=
-\frac{r_0^2}{r^2},
\qquad
b'(r_0)=-1<1,
\label{eq:flareout_explicit}
\end{equation}
so that Equation~\eqref{eq:flareout_condition} is~satisfied.

The radial metric coefficient follows directly from
Equations~\eqref{eq:ads_teo_metric}, \eqref{eq:metric_functions}, and~\eqref{eq:A_definition}:
\begin{equation}
g_{rr}
=
\frac{1}
{
\left(1-\dfrac{r_0^2}{r^2}\right)
\left(1+\dfrac{r^2-r_0^2}{L^2}\right)
}.
\label{eq:grr_exact}
\end{equation}

For $r\rightarrow\infty$,
\begin{equation}
\mathcal{A}(r)
=
\frac{r^2}{L^2}
+
\left(
1-\frac{r_0^2}{L^2}
\right),
\label{eq:A_asymptotic}
\end{equation}
and, therefore,
\begin{equation}
g_{rr}
=
\frac{L^2}{r^2}
+
\mathcal{O}(r^{-4}),
\qquad
r\rightarrow\infty.
\label{eq:grr_ads_asymptotic}
\end{equation}

Similarly,
\begin{equation}
N_0^2(r)
=
\frac{r^2}{L^2}
+
\left(
1-\frac{r_0^2}{L^2}
\right),
\label{eq:N0_ads_asymptotic}
\end{equation}
so the leading temporal and radial sectors have the required AdS
behavior.  The~frame-dragging term satisfies
\begin{equation}
\Omega(r)
=
\frac{2a}{r^3}
=
\mathcal{O}(r^{-3}),
\label{eq:Omega_asymptotic}
\end{equation}
and hence does not modify the leading asymptotic AdS~structure.

To generate genuine non-separability while preserving both the throat
and the asymptotic geometry, we introduce the axisymmetric quadrupolar
lapse deformation
\begin{equation}
\mathcal{D}(r,\theta)
=
\epsilon\,q(r)\,P_2(\cos\theta),
\label{eq:lapse_deformation}
\end{equation}
where
\begin{equation}
P_2(\cos\theta)
=
\frac{1}{2}
\left(
3\cos^2\theta-1
\right)
\label{eq:P2_definition}
\end{equation}
and
\begin{equation}
q(r)
=
\frac{
8r_0^2\left(r^2-r_0^2\right)
}{
\left(r^2+r_0^2\right)^2
}.
\label{eq:q_profile}
\end{equation}

The parameter $\epsilon$ controls the deformation strength.  The~radial
profile satisfies
\begin{equation}
q(r_0)=0,
\qquad
q(r)
=
\frac{8r_0^2}{r^2}
+
\mathcal{O}(r^{-4}),
\qquad
r\rightarrow\infty,
\label{eq:q_limits}
\end{equation}
and reaches unit amplitude at
\begin{equation}
q(\sqrt{3}\,r_0)=1.
\label{eq:q_normalization}
\end{equation}

Thus, $\epsilon$ measures directly the maximum radial amplitude of the
deformation.  Since $P_2(\cos\theta)$ is independent of $\phi$ and even
under $\theta\mapsto\pi-\theta$, Equation~\eqref{eq:lapse_deformation}
preserves stationarity, axisymmetry, and~equatorial reflection symmetry.
Because Equation~\eqref{eq:q_limits} makes the deformation vanish at the
throat and at both asymptotic ends, the~canonical throat geometry and
the leading AdS asymptotics are unchanged. The~lapse, therefore, has \mbox{the limits}
\begin{equation}
N(r_0,\theta)=1,
\label{eq:lapse_throat}
\end{equation}
and
\begin{equation}
N(r,\theta)
=
\frac{r}{L}
+
\mathcal{O}(r^{-1}),
\qquad
r\rightarrow\infty,
\label{eq:lapse_asymptotic}
\end{equation}
where the angular deformation first contributes at subleading order.
The geometry consequently interpolates between a regular rotating Teo
throat and two asymptotically AdS regions while retaining an explicit
$\theta$ dependence in the intermediate radial~domain.

For later use, the~nonvanishing metric components in the
$(t,\phi)$ sector are
\begin{align}
g_{tt}
&=
-N^2+r^2\sin^2\theta\,\Omega^2,
\label{eq:gtt_exact}
\\
g_{t\phi}
&=
-r^2\sin^2\theta\,\Omega,
\label{eq:gtphi_exact}
\\
g_{\phi\phi}
&=
r^2\sin^2\theta.
\label{eq:gphiphi_exact}
\end{align}

Their determinant is
\begin{equation}
g_{tt}g_{\phi\phi}-g_{t\phi}^2
=
-N^2r^2\sin^2\theta,
\label{eq:tphi_determinant}
\end{equation}
which gives the inverse components
\begin{align}
g^{tt}
&=
-\frac{1}{N^2},
\label{eq:gtt_inverse}
\\
g^{t\phi}
&=
-\frac{\Omega}{N^2},
\label{eq:gtphi_inverse}
\\
g^{\phi\phi}
&=
\frac{1}{r^2\sin^2\theta}
-
\frac{\Omega^2}{N^2}.
\label{eq:gphiphi_inverse}
\end{align}

The remaining inverse components are
\begin{equation}
g^{rr}
=
\left(1-\frac{b}{r}\right)\mathcal{A},
\qquad
g^{\theta\theta}
=
\frac{1}{r^2},
\label{eq:inverse_radial_angular}
\end{equation}
and the metric determinant is
\begin{equation}
\sqrt{-g}
=
\frac{
Nr^2\sin\theta
}{
\sqrt{
\left(1-\dfrac{b}{r}\right)\mathcal{A}
}
}.
\label{eq:metric_determinant}
\end{equation}

Before formulating the spectral problem, the~admissible rotational
parameter range must be identified.  An~ergoregion occurs wherever the
stationary Killing vector $\xi=\partial_t$ becomes spacelike,
\begin{equation}
g_{tt}>0.
\label{eq:ergoregion_definition}
\end{equation}

Using Equations~\eqref{eq:gtt_exact} and \eqref{eq:metric_functions}, the~ergosurface condition is
\begin{equation}
N^2(r,\theta)
=
\frac{4a^2}{r^4}\sin^2\theta.
\label{eq:ergosurface_condition}
\end{equation}

Hence a sufficient condition for the absence of an ergoregion is
\begin{equation}
N(r,\theta)
>
\frac{2|a|}{r^2}
\label{eq:ergoregion_sufficient_local}
\end{equation}
throughout the spacetime. Since
\begin{equation}
-\frac{1}{2}
\leq
P_2(\cos\theta)
\leq
1
\label{eq:P2_bounds}
\end{equation}
and the normalized profile obeys $0\leq q(r)\leq1$, the~lapse satisfies,
for $\epsilon\geq0$,
\begin{equation}
N(r,\theta)
=
N_0(r)+\epsilon q(r)P_2(\cos\theta)
\geq
N_0(r)-\frac{\epsilon}{2}q(r)
\geq
N_0(r)-\frac{\epsilon}{2}.
\label{eq:lapse_lower_bound}
\end{equation}

Because $N_0(r)\geq N_0(r_0)=1$ for $r\geq r_0$,
\begin{equation}
N(r,\theta)
\geq
1-\frac{\epsilon}{2}.
\label{eq:lapse_global_lower_bound}
\end{equation}

At the same time,
\begin{equation}
\frac{2|a|}{r^2}
\leq
\frac{2|a|}{r_0^2}.
\label{eq:rotation_global_upper_bound}
\end{equation}

Combining Equations~\eqref{eq:ergoregion_sufficient_local},
\eqref{eq:lapse_global_lower_bound}, and~\eqref{eq:rotation_global_upper_bound} gives the sufficient {global bound}

\begin{equation}
\boxed{
|a|
<
\frac{r_0^2}{2}
\left(
1-\frac{\epsilon}{2}
\right).
}
\label{eq:ergoregion_global_bound}
\end{equation}

All spectral and quantum calculations below are restricted to
Equation~\eqref{eq:ergoregion_global_bound}. Within~this domain the
stationary Killing vector remains timelike everywhere. Outside this
sufficiently restricted domain, Equation~\eqref{eq:ergoregion_global_bound}
alone no longer guarantees the absence of an ergoregion, and~no
stability conclusion is drawn~here.

The throat coordinate singularity in $r$ is removed by introducing the
proper radial distance $\ell$.  Along curves of fixed
$(t,\theta,\phi)$, Equation~\eqref{eq:ads_teo_metric} gives
\begin{equation}
d\ell^2
=
\frac{dr^2}
{
\left(1-\dfrac{b(r)}{r}\right)\mathcal{A}(r)
},
\label{eq:proper_distance_line}
\end{equation}
so that
\begin{equation}
\frac{d\ell}{dr}
=
\pm
\frac{1}{
\sqrt{
\left(1-\dfrac{b(r)}{r}\right)\mathcal{A}(r)
}
}.
\label{eq:proper_distance_definition}
\end{equation}

The two signs describe the two asymptotic regions.  We choose
\begin{equation}
\ell(r_0)=0,
\qquad
-\infty<\ell<\infty,
\label{eq:ell_domain}
\end{equation}
with $\ell\rightarrow\pm\infty$ approaching the two AdS~boundaries.

For the explicit functions in
Equations~\eqref{eq:metric_functions} and \eqref{eq:A_definition},
Equation~\eqref{eq:proper_distance_definition} becomes
\begin{equation}
\frac{d\ell}{dr}
=
\pm
\frac{r}{
\sqrt{
\left(r^2-r_0^2\right)
\left[
1+\dfrac{r^2-r_0^2}{L^2}
\right]
}
}.
\label{eq:proper_distance_explicit}
\end{equation}

Introducing
\begin{equation}
u=\sqrt{r^2-r_0^2},
\qquad
du=\frac{r\,dr}{\sqrt{r^2-r_0^2}},
\label{eq:proper_distance_substitution}
\end{equation}
gives
\begin{equation}
d\ell
=
\pm
\frac{du}{\sqrt{1+u^2/L^2}},
\label{eq:proper_distance_integral}
\end{equation}
and hence
\begin{equation}
\ell
=
\pm L\,
\sinh^{-1}
\left(
\frac{\sqrt{r^2-r_0^2}}{L}
\right).
\label{eq:ell_of_r}
\end{equation}

Inverting,
\begin{equation}
\boxed{
r^2(\ell)
=
r_0^2
+
L^2\sinh^2\left(\frac{\ell}{L}\right).
}
\label{eq:r_of_ell}
\end{equation}

It follows that
\begin{equation}
\mathcal{A}(\ell)
=
\cosh^2\left(\frac{\ell}{L}\right),
\qquad
N_0(\ell)
=
\cosh\left(\frac{\ell}{L}\right),
\label{eq:A_N_ell}
\end{equation}
and
\begin{equation}
\frac{dr}{d\ell}
=
\frac{
L\sinh(\ell/L)\cosh(\ell/L)
}{
r(\ell)
}.
\label{eq:dr_dell}
\end{equation}

Near the throat,
\begin{equation}
r(\ell)
=
r_0
+
\frac{\ell^2}{2r_0}
+
\mathcal{O}(\ell^4),
\label{eq:r_near_throat}
\end{equation}
whereas for $|\ell|\rightarrow\infty$,
\begin{equation}
r(\ell)
\sim
\frac{L}{2}
e^{|\ell|/L}.
\label{eq:r_asymptotic_ell}
\end{equation}

Thus, $\ell$ covers both asymptotic regions smoothly and places the
wormhole throat at the regular interior point $\ell=0$.


In the proper-distance coordinate it is convenient to define the
angular areal factor
\begin{equation}
\rho(\ell,\theta)
\equiv
r(\ell)K(\ell,\theta).
\label{eq:rho_definition}
\end{equation}

For the background specified in Equation~\eqref{eq:metric_functions},
$K=1$ and, therefore,
\begin{equation}
\rho(\ell,\theta)=r(\ell).
\label{eq:rho_equals_r}
\end{equation}

Equation~\eqref{eq:ads_teo_metric} then becomes
\begin{equation}
\begin{aligned}
ds^2={}&
-N^2(\ell,\theta)\,dt^2
+d\ell^2
\\
&+
\rho^2(\ell,\theta)
\left[
d\theta^2
+
\sin^2\theta
\left(
d\phi-\Omega(\ell,\theta)\,dt
\right)^2
\right].
\end{aligned}
\label{eq:proper_distance_metric}
\end{equation}

Reflection symmetry of the wormhole gives
\begin{equation}
N(-\ell,\theta)=N(\ell,\theta),
\qquad
\rho(-\ell,\theta)=\rho(\ell,\theta),
\qquad
\Omega(-\ell,\theta)=\Omega(\ell,\theta).
\label{eq:reflection_symmetry_metric}
\end{equation}

The transformation $\ell\mapsto-\ell$ exchanges the two asymptotic AdS
regions while leaving the geometry~invariant.

We now introduce a minimally coupled probe scalar field of mass $\mu$ \cite{WaldQFTCS},
\begin{equation}
\left(
\Box-\mu^2
\right)\Phi=0,
\label{eq:KG_equation}
\end{equation}
with
\begin{equation}
\Box\Phi
=
\frac{1}{\sqrt{-g}}
\partial_\mu
\left(
\sqrt{-g}\,
g^{\mu\nu}
\partial_\nu\Phi
\right).
\label{eq:dAlembertian_definition}
\end{equation}

Stationarity and axisymmetry allow the exact Fourier decomposition
\begin{equation}
\Phi(t,\ell,\theta,\phi)
=
e^{-i\omega t}
e^{im\phi}
\Psi_{\omega m}(\ell,\theta),
\label{eq:Fourier_scalar}
\end{equation}

Throughout the fixed-$m$ reduction below, $Y_{Lm}(\theta)$ denotes the
normalized polar harmonic obtained after extracting the common
azimuthal dependence $e^{im\phi}$, with~normalization given in
Equation~\eqref{eq:harmonic_orthogonality_ads}. Full conventionally
normalized spherical harmonics $Y_{Lm}(\theta,\phi)$ are restored
explicitly whenever angular integrals are written over $d\Omega$.
No further factorization
$\Psi_{\omega m}(\ell,\theta)=R(\ell)S(\theta)$ is assumed. Such a
separation requires additional hidden symmetry of the type present in
Kerr spacetime~\cite{Carter1968,Teukolsky1973,Chandrasekhar1983}; no
analogous rank-two Killing tensor is assumed for the deformed geometry
defined by Equation~\eqref{eq:lapse_deformation}.

Expanding the square in Equation~\eqref{eq:proper_distance_metric} gives
\begin{equation}
\begin{aligned}
ds^2={}&
\left(
-N^2+\rho^2\Omega^2\sin^2\theta
\right)dt^2
-2\rho^2\Omega\sin^2\theta\,dt\,d\phi
\\
&+
d\ell^2
+\rho^2d\theta^2
+\rho^2\sin^2\theta\,d\phi^2 .
\end{aligned}
\label{eq:proper_metric_expanded}
\end{equation}

Hence
\begin{align}
g_{tt}
&=
-N^2+\rho^2\Omega^2\sin^2\theta,
\label{eq:gtt_ell}
\\
g_{t\phi}
&=
-\rho^2\Omega\sin^2\theta,
\label{eq:gtphi_ell}
\\
g_{\phi\phi}
&=
\rho^2\sin^2\theta,
\label{eq:gphiphi_ell}
\\
g_{\ell\ell}
&=1,
\qquad
g_{\theta\theta}=\rho^2.
\label{eq:gell_gtheta}
\end{align}

The determinant of the $(t,\phi)$ block is
\begin{equation}
g_{tt}g_{\phi\phi}-g_{t\phi}^{\,2}
=
-N^2\rho^2\sin^2\theta,
\label{eq:tphi_block_det_ell}
\end{equation}
and, therefore,
\begin{align}
g^{tt}
&=
-\frac{1}{N^2},
\label{eq:gtt_inv_ell}
\\
g^{t\phi}
&=
-\frac{\Omega}{N^2},
\label{eq:gtphi_inv_ell}
\\
g^{\phi\phi}
&=
\frac{1}{\rho^2\sin^2\theta}
-\frac{\Omega^2}{N^2},
\label{eq:gphiphi_inv_ell}
\\
g^{\ell\ell}
&=1,
\qquad
g^{\theta\theta}=\frac{1}{\rho^2}.
\label{eq:gell_gtheta_inv}
\end{align}

The full determinant is
\begin{equation}
\sqrt{-g}
=
N\rho^2\sin\theta.
\label{eq:sqrt_minus_g_ell}
\end{equation}

Substituting Equation~\eqref{eq:Fourier_scalar} into
Equation~\eqref{eq:KG_equation}, the~temporal and azimuthal derivatives
become algebraic while the $\ell$ and $\theta$ derivatives remain
differential.  Before~simplifying the rotating sector one obtains
\begin{equation}
\begin{aligned}
0={}&
\frac{1}{\sqrt{-g}}
\partial_\ell
\left(
\sqrt{-g}\,
\partial_\ell\Psi_{\omega m}
\right)
+
\frac{1}{\sqrt{-g}}
\partial_\theta
\left(
\sqrt{-g}\,
g^{\theta\theta}
\partial_\theta\Psi_{\omega m}
\right)
\\
&+
\left(
-g^{tt}\omega^2
+
2g^{t\phi}\omega m
-
g^{\phi\phi}m^2
-\mu^2
\right)
\Psi_{\omega m}.
\end{aligned}
\label{eq:KG_intermediate}
\end{equation}

Using Equations~\eqref{eq:gtt_inv_ell}--\eqref{eq:gphiphi_inv_ell},
\begin{equation}
-g^{tt}\omega^2
+
2g^{t\phi}\omega m
-
g^{\phi\phi}m^2
=
\frac{(\omega-m\Omega)^2}{N^2}
-
\frac{m^2}{\rho^2\sin^2\theta}.
\label{eq:corotating_combination}
\end{equation}

Together with Equation~\eqref{eq:sqrt_minus_g_ell}, the~scalar equation is,
therefore,
\begin{equation}
\boxed{
\begin{aligned}
0={}&
\frac{1}{N\rho^2\sin\theta}
\partial_\ell
\left(
N\rho^2\sin\theta\,
\partial_\ell\Psi_{\omega m}
\right)
\\
&+
\frac{1}{N\rho^2\sin\theta}
\partial_\theta
\left(
N\sin\theta\,
\partial_\theta\Psi_{\omega m}
\right)
\\
&+
\left[
\frac{(\omega-m\Omega)^2}{N^2}
-
\frac{m^2}{\rho^2\sin^2\theta}
-\mu^2
\right]
\Psi_{\omega m}.
\end{aligned}
}
\label{eq:full_scalar_equation}
\end{equation}

The explicit dependence
\[
N(\ell,\theta)
=
N_0(\ell)
+
\epsilon q(\ell)P_2(\cos\theta)
\]
from Equation~\eqref{eq:lapse_deformation} prevents the angular sector of
Equation~\eqref{eq:full_scalar_equation} from being diagonalized into an
independent radial~equation.

The angular dependence is instead expanded at fixed $m$ in the complete
spherical-harmonic basis,
\begin{equation}
\Psi_{\omega m}(\ell,\theta)
=
\sum_{L\geq |m|}
R_L(\ell)Y_{Lm}(\theta),
\label{eq:harmonic_expansion_ads}
\end{equation}
with normalization
\begin{equation}
\int_0^\pi
d\theta\,
\sin\theta\,
Y_{Lm}(\theta)
Y_{L'm}(\theta)
=
\delta_{LL'}.
\label{eq:harmonic_orthogonality_ads}
\end{equation}

Multiplying Equation~\eqref{eq:full_scalar_equation} by
$N\rho^2\sin\theta$ gives
\begin{equation}
\begin{aligned}
0={}&
\partial_\ell
\left(
N\rho^2\sin\theta\,
\partial_\ell\Psi_{\omega m}
\right)
+
\partial_\theta
\left(
N\sin\theta\,
\partial_\theta\Psi_{\omega m}
\right)
\\
&+
\sin\theta
\left[
\frac{\rho^2}{N}
(\omega-m\Omega)^2
-
\frac{Nm^2}{\sin^2\theta}
-
N\rho^2\mu^2
\right]
\Psi_{\omega m}.
\end{aligned}
\label{eq:KG_weighted}
\end{equation}

Substitution of Equation~\eqref{eq:harmonic_expansion_ads} gives
\begin{equation}
\partial_\ell\Psi_{\omega m}
=
\sum_{L'}
\frac{dR_{L'}}{d\ell}
Y_{L'm},
\qquad
\partial_\theta\Psi_{\omega m}
=
\sum_{L'}
R_{L'}
\partial_\theta Y_{L'm}.
\label{eq:harmonic_derivatives}
\end{equation}

Projection onto $Y_{Lm}$ then yields
\begin{equation}
\sum_{L'}
\left[
\frac{d}{d\ell}
\left(
P_{LL'}(\ell)
\frac{dR_{L'}}{d\ell}
\right)
+
Q_{LL'}(\ell;\omega,m)
R_{L'}(\ell)
\right]
=0,
\label{eq:projected_matrix_system}
\end{equation}
where the kinetic matrix is
\begin{equation}
P_{LL'}(\ell)
=
\int_0^\pi
d\theta\,
\sin\theta\,
Y_{Lm}(\theta)
N(\ell,\theta)
\rho^2(\ell,\theta)
Y_{L'm}(\theta)
\label{eq:P_matrix_ads}
\end{equation}
and
\begin{equation}
\begin{aligned}
Q_{LL'}(\ell;\omega,m)
={}&
\int_0^\pi
d\theta\,
Y_{Lm}
\partial_\theta
\left[
N\sin\theta\,
\partial_\theta Y_{L'm}
\right]
\\
&+
\int_0^\pi
d\theta\,
\sin\theta\,
Y_{Lm}
\Bigg[
\frac{\rho^2}{N}
(\omega-m\Omega)^2
\\
&\hspace{30mm}
-
\frac{Nm^2}{\sin^2\theta}
-
N\rho^2\mu^2
\Bigg]
Y_{L'm}.
\end{aligned}
\label{eq:Q_matrix_ads}
\end{equation}

The radial divergence in
Equation~\eqref{eq:projected_matrix_system} acts on the complete product,
\begin{equation}
\frac{d}{d\ell}
\left(
P_{LL'}
\frac{dR_{L'}}{d\ell}
\right)
=
P_{LL'}
\frac{d^2R_{L'}}{d\ell^2}
+
\frac{dP_{LL'}}{d\ell}
\frac{dR_{L'}}{d\ell}.
\label{eq:P_derivative_expansion}
\end{equation}

The second term in Equation~\eqref{eq:P_derivative_expansion} is generated
by the radial variation of the geometry and is retained throughout the
spectral~analysis.

The explicit frequency dependence follows from
\begin{equation}
\frac{\rho^2}{N}
(\omega-m\Omega)^2
=
\omega^2\frac{\rho^2}{N}
-
2m\omega\frac{\rho^2\Omega}{N}
+
m^2\frac{\rho^2\Omega^2}{N}.
\label{eq:frequency_expansion_ads}
\end{equation}

Accordingly,
\begin{equation}
Q_{LL'}(\ell;\omega,m)
=
\omega^2W_{LL'}(\ell)
+
\omega C_{LL'}(\ell)
+
H_{LL'}(\ell),
\label{eq:Q_decomposition_ads}
\end{equation}
with
\begin{equation}
W_{LL'}(\ell)
=
\int_0^\pi
d\theta\,
\sin\theta\,
Y_{Lm}
\frac{\rho^2}{N}
Y_{L'm},
\label{eq:W_matrix_ads}
\end{equation}
\begin{equation}
C_{LL'}(\ell)
=
-2m
\int_0^\pi
d\theta\,
\sin\theta\,
Y_{Lm}
\frac{\rho^2\Omega}{N}
Y_{L'm},
\label{eq:C_matrix_ads}
\end{equation}
and
\begin{equation}
\begin{aligned}
H_{LL'}(\ell)
={}&
\int_0^\pi
d\theta\,
Y_{Lm}
\partial_\theta
\left[
N\sin\theta\,
\partial_\theta Y_{L'm}
\right]
\\
&+
\int_0^\pi
d\theta\,
\sin\theta\,
Y_{Lm}
\Bigg[
m^2\frac{\rho^2\Omega^2}{N}
-
\frac{Nm^2}{\sin^2\theta}
-
N\rho^2\mu^2
\Bigg]
Y_{L'm}.
\end{aligned}
\label{eq:H_matrix_ads}
\end{equation}

The geometry-derived coupled radial equations are, therefore,
\begin{equation}
\boxed{
\sum_{L'}
\left\{
\frac{d}{d\ell}
\left[
P_{LL'}(\ell)
\frac{dR_{L'}}{d\ell}
\right]
+
\left[
\omega^2W_{LL'}(\ell)
+
\omega C_{LL'}(\ell)
+
H_{LL'}(\ell)
\right]
R_{L'}(\ell)
\right\}
=0.
}
\label{eq:coupled_radial_PWCH}
\end{equation}

Thus, the metric fixes not only the static interaction matrix but also the
kinetic matrix and both frequency-dependent matrices.  No independent
phenomenological channel-coupling term is~introduced.

In a fixed angular-parity sector, the~channel vector may be written as
\begin{equation}
\mathbf R(\ell)
=
\left(
R_{|m|},
R_{|m|+2},
R_{|m|+4},
\ldots
\right)^{T},
\label{eq:channel_vector}
\end{equation}
and Equation~\eqref{eq:coupled_radial_PWCH} becomes
\begin{equation}
\frac{d}{d\ell}
\left(
\mathbf P(\ell)
\frac{d\mathbf R}{d\ell}
\right)
+
\left[
\omega^2\mathbf W(\ell)
+
\omega\mathbf C(\ell)
+
\mathbf H(\ell)
\right]
\mathbf R(\ell)
=
0.
\label{eq:matrix_radial_compact}
\end{equation}

The diagonal entries describe propagation within an angular channel,
whereas the off-diagonal entries describe channel mixing generated by
the explicit angular dependence of the background~geometry.

The angular derivative contribution in
Equation~\eqref{eq:H_matrix_ads} may also be placed in manifestly symmetric
form.  Integration by parts gives
\begin{equation}
\begin{aligned}
&\int_0^\pi
d\theta\,
Y_{Lm}
\partial_\theta
\left[
N\sin\theta\,
\partial_\theta Y_{L'm}
\right]
\\
&\qquad
=
-
\int_0^\pi
d\theta\,
N\sin\theta\,
(\partial_\theta Y_{Lm})
(\partial_\theta Y_{L'm}),
\end{aligned}
\label{eq:angular_integration_by_parts}
\end{equation}
because the boundary contribution vanishes at
$\theta=0,\pi$.  The~symmetry of the angular projection is, therefore,
explicit.

\textls[-25]{The quadrupolar origin of the off-diagonal structure follows directly
from \mbox{Equation~\eqref{eq:lapse_deformation}.}}  The~relevant first-order overlap
matrix is
\begin{equation}
\mathcal{M}^{(m)}_{LL'}
=
\int d\Omega\,
Y_{Lm}^{*}(\theta,\phi)
P_2(\cos\theta)
Y_{L'm}(\theta,\phi).
\label{eq:P2_overlap_general}
\end{equation}

Since
\begin{equation}
P_2(\cos\theta)
=
\sqrt{\frac{4\pi}{5}}\,
Y_{20}(\theta,\phi),
\label{eq:P2_Y20}
\end{equation}

Equation~\eqref{eq:P2_overlap_general} is a Gaunt integral.  Using
\begin{equation}
Y_{Lm}^{*}
=
(-1)^mY_{L,-m},
\label{eq:Y_conjugation}
\end{equation}
one obtains
\begin{equation}
\mathcal{M}^{(m)}_{LL'}
=
(-1)^m
\sqrt{(2L+1)(2L'+1)}
\begin{pmatrix}
L & 2 & L'\\
0 & 0 & 0
\end{pmatrix}
\begin{pmatrix}
L & 2 & L'\\
-m & 0 & m
\end{pmatrix}.
\label{eq:Gaunt_overlap_ads}
\end{equation}

The Wigner-symbol conditions imply
\begin{equation}
|L-L'|\leq2\leq L+L',
\qquad
L+L'+2=\text{even},
\label{eq:Wigner_selection_conditions}
\end{equation}
and, therefore,
\begin{equation}
\boxed{
L'=L,\;L\pm2.
}
\label{eq:selection_rule_ads}
\end{equation}

The deformation consequently preserves angular parity while coupling
neighboring same-parity channels separated by $\Delta L=2$.

For the two-channel sector used later in the explicit geometry-derived
reduction,
\begin{equation}
m=1,
\qquad
L\in\{1,3\},
\label{eq:two_channel_sector_ads}
\end{equation}
the channel vector is
\begin{equation}
\mathbf R(\ell)
=
\begin{pmatrix}
R_1(\ell)\\
R_3(\ell)
\end{pmatrix}.
\label{eq:R13_vector}
\end{equation}

Higher truncations follow from the same projection by enlarging the
retained basis according to
\begin{equation}
L
=
|m|,
|m|+2,
|m|+4,
\ldots .
\label{eq:channel_hierarchy_ads}
\end{equation}

The accuracy of these finite-dimensional projections is tested
explicitly through the two-, four-, and~six-channel convergence
calculations~below.

Finally, the~proper-distance coordinate makes the local throat
structure regular.  Since the geometry is invariant under
$\ell\mapsto-\ell$, every background coefficient has an even \mbox{expansion,}
\begin{equation}
F(\ell,\theta)
=
F_0(\theta)
+
F_2(\theta)\ell^2
+
\mathcal{O}(\ell^4),
\label{eq:even_throat_expansion}
\end{equation}
for
$F=N,\rho,\Omega$. Using the near-throat expansion in
Equation~\eqref{eq:r_near_throat},
\begin{equation}
\mathbf P(\ell)
=
\mathbf P^{(0)}
+
\mathbf P^{(2)}\ell^2
+
\mathcal{O}(\ell^4),
\qquad
\mathbf Q(\ell;\omega)
=
\mathbf Q^{(0)}(\omega)
+
\mathbf Q^{(2)}(\omega)\ell^2
+
\mathcal{O}(\ell^4),
\label{eq:PQ_throat_expansion}
\end{equation}
and Equation~\eqref{eq:projected_matrix_system} reduces at leading order to
\begin{equation}
\sum_{L'}
\left[
P^{(0)}_{LL'}R''_{L'}
+
Q^{(0)}_{LL'}R_{L'}
\right]
=
\begin{cases}
\mathcal O(\ell^2), & \text{even sector},\\[1mm]
\mathcal O(\ell),   & \text{odd sector}.
\end{cases}
\label{eq:near_throat_matrix_system}
\end{equation}

The throat is, therefore, a smooth interior point rather than a horizon.
Regular solutions may be classified by parity,
\begin{equation}
\left.
\frac{dR_L}{d\ell}
\right|_{\ell=0}
=0
\qquad
\text{(even sector)},
\label{eq:even_throat_bc}
\end{equation}
or
\begin{equation}
R_L(0)=0
\qquad
\text{(odd sector)}.
\label{eq:odd_throat_bc}
\end{equation}

No ingoing or outgoing condition is imposed at $\ell=0$.  The~discrete
normal-mode spectrum is obtained only after these regular throat
solutions are combined with the normalizable boundary conditions at the
two asymptotic AdS~ends.

\section{Global Coupled Spectral Problem and Normal~Modes}
\label{sec:global_spectral}

The coupled equations derived in
Equation~\eqref{eq:matrix_radial_compact} define a boundary-value problem on
the complete wormhole domain $-\infty<\ell<\infty$.  Reflection
symmetry, Equation~\eqref{eq:reflection_symmetry_metric}, acts independently
of the angular harmonic labels and allows the radial channel amplitudes
to be classified according to
\begin{equation}
R_L(-\ell)
=
\pm R_L(\ell),
\label{eq:radial_parity}
\end{equation}
with the upper and lower signs denoting the even and odd sectors,
respectively.  Hence a parity eigenmode satisfies
\begin{equation}
\int_{-\infty}^{+\infty}
d\ell\,
\mathbf R^\dagger
\mathbf P
\mathbf R
=
2
\int_0^\infty
d\ell\,
\mathbf R^\dagger
\mathbf P
\mathbf R,
\label{eq:parity_halfline_integral}
\end{equation}
because $\mathbf P(-\ell)=\mathbf P(\ell)$.  The~full two-sided problem
may, therefore, be represented on the half-line $\ell\geq0$ with the
throat conditions already obtained in
Equations~\eqref{eq:even_throat_bc} and \eqref{eq:odd_throat_bc}.

For a truncation containing $N$ angular channels, we take the underlying
Hilbert space to be
\begin{equation}
\mathcal H_N
=
L^2
\left(
(0,\infty),\mathbb C^N
\right),
\label{eq:finite_Hilbert_space}
\end{equation}
with inner product
\begin{equation}
\left(
\mathbf R_1,\mathbf R_2
\right)
=
\int_0^\infty
d\ell\,
\mathbf R_1^\dagger(\ell)
\mathbf R_2(\ell).
\label{eq:standard_inner_product}
\end{equation}

Formally, the~infinite-channel limit is obtained by replacing
$\mathbb C^N$ by the sequence \mbox{space $\ell^2$,}
\begin{equation}
\mathcal H
=
L^2
\left(
(0,\infty),\ell^2
\right).
\label{eq:infinite_Hilbert_space}
\end{equation}

The operator domain consists of sufficiently regular channel vectors
for which the differential expression in
Equation~\eqref{eq:matrix_radial_compact} is well defined and which satisfy
one of the parity conditions at $\ell=0$ together with the chosen
normalizable AdS boundary condition at infinity.  For~a finite
truncation we write
\begin{equation}
\begin{aligned}
D(\mathcal L_\pm)
=
\Bigl\{
\mathbf R\in\mathcal H_N:\;&
\mathbf R,\,
\mathbf P\mathbf R'
\ \text{locally absolutely continuous},
\\
&
\mathcal L\mathbf R\in\mathcal H_N,
\quad
\mathcal B_\pm[\mathbf R]=0,
\\
&
\mathcal B_{\rm AdS}[\mathbf R]=0
\Bigr\}.
\end{aligned}
\label{eq:operator_domain}
\end{equation}
where
\begin{equation}
\mathcal B_+[\mathbf R]
=
\mathbf R'(0),
\qquad
\mathcal B_-[\mathbf R]
=
\mathbf R(0)
\label{eq:parity_boundary_operators}
\end{equation}
represent the even and odd throat~conditions.

\textls[-25]{The frequency-independent differential operator is defined directly
from Equation~\eqref{eq:matrix_radial_compact} by}
\begin{equation}
\mathcal A
=
-
\frac{d}{d\ell}
\left(
\mathbf P(\ell)
\frac{d}{d\ell}
\right)
-
\mathbf H(\ell).
\label{eq:A_operator}
\end{equation}

The rotating problem is then governed by the quadratic operator pencil
\begin{equation}
\boxed{
\mathcal P(\omega)
=
\mathcal A
-
\omega\mathbf C(\ell)
-
\omega^2\mathbf W(\ell),
}
\label{eq:quadratic_operator_pencil}
\end{equation}
and the normal-mode problem is
\begin{equation}
\mathcal P(\omega)\mathbf R=0.
\label{eq:pencil_eigenproblem}
\end{equation}

Thus, the matrix Sturm--Liouville differential expression
$\mathcal A$ and the frequency-dependent spectral pencil
$\mathcal P(\omega)$ are distinct mathematical objects.  The~former
defines the underlying differential operator, whereas the latter
determines the allowed spectral values of $\omega$.

In addition to the standard Hilbert-space inner product
\eqref{eq:standard_inner_product}, the~kinetic matrix defines the
weighted bilinear form
\begin{equation}
\left\langle
\mathbf R_1,\mathbf R_2
\right\rangle_{\mathbf P}
=
\int_0^\infty
d\ell\,
\mathbf R_1^\dagger(\ell)
\mathbf P(\ell)
\mathbf R_2(\ell).
\label{eq:P_weighted_inner_product}
\end{equation}

Integration by parts gives
\begin{equation}
\begin{aligned}
&
\left(
\mathbf R_1,
\mathcal A\mathbf R_2
\right)
-
\left(
\mathcal A\mathbf R_1,
\mathbf R_2
\right)
\\
&\qquad
=
\left[
\mathbf R_1^\dagger
\mathbf P
\frac{d\mathbf R_2}{d\ell}
-
\left(
\frac{d\mathbf R_1}{d\ell}
\right)^\dagger
\mathbf P
\mathbf R_2
\right]_{0}^{\infty}.
\end{aligned}
\label{eq:matrix_boundary_form}
\end{equation}

The throat contribution vanishes separately in either parity sector,
while the asymptotic term vanishes for the normalizable AdS boundary
conditions imposed below.  Consequently, $\mathcal A$ is symmetric on
the domain \eqref{eq:operator_domain} and admits the corresponding
self-adjoint realization familiar from matrix Sturm--Liouville theory
~\cite{Zettl2005,Teschl2014}.

This statement concerns the differential operator $\mathcal A$.
Because rotation introduces the linear and quadratic frequency terms in
Equation~\eqref{eq:quadratic_operator_pencil}, the~complete spectral problem
must still be treated as a quadratic operator-pencil eigenvalue
problem.  Throughout the ergoregion-free parameter regime
\eqref{eq:ergoregion_global_bound}, together with the conservative
throat and reflecting AdS boundary conditions, we, therefore, formulate
the problem as a conservative normal-mode eigenvalue problem.  The~complete numerical eigenspectrum studied below resolves no nonzero
imaginary components within the numerical precision of the
calculations over the parameter range considered.
Near the throat, each channel admits a regular Taylor expansion
\begin{equation}
R_L(\ell)
=
R_{L,0}
+
R_{L,1}\ell
+
R_{L,2}\ell^2
+
\cdots .
\label{eq:throat_Taylor_modes}
\end{equation}

The parity conditions restrict the expansion to
\begin{equation}
R_L^{(+)}(\ell)
=
R_{L,0}
+
R_{L,2}\ell^2
+
R_{L,4}\ell^4
+\cdots
\label{eq:even_Taylor_modes}
\end{equation}
or
\begin{equation}
R_L^{(-)}(\ell)
=
R_{L,1}\ell
+
R_{L,3}\ell^3
+
R_{L,5}\ell^5
+\cdots .
\label{eq:odd_Taylor_modes}
\end{equation}

These expansions provide an $N$-dimensional set of regular initial data
within each parity~sector.

We next determine the allowed asymptotic behavior.  Using the
large-$r$ geometry established in
Equations~\eqref{eq:grr_ads_asymptotic} and
\eqref{eq:N0_ads_asymptotic}, the~leading coefficients entering the
Klein--Gordon equation are
\begin{equation}
\sqrt{-g}
\sim
r^2\sin\theta,
\qquad
g^{rr}
\sim
\frac{r^2}{L^2}.
\label{eq:asymptotic_metric_coefficients}
\end{equation}

The frequency, angular, and~frame-dragging terms are subleading in the
large-$r$ indicial equation, so the leading Klein--Gordon equation
reduces to
\begin{equation}
\frac{1}{r^2}
\frac{d}{dr}
\left(
\frac{r^4}{L^2}
\frac{d\Psi}{dr}
\right)
-
\mu^2\Psi
\simeq
0.
\label{eq:asymptotic_KG_radial}
\end{equation}

Multiplying by $L^2$ gives
\begin{equation}
r^2
\frac{d^2\Psi}{dr^2}
+
4r
\frac{d\Psi}{dr}
-
\mu^2L^2\Psi
\simeq
0.
\label{eq:asymptotic_Euler_equation}
\end{equation}

Taking
\begin{equation}
\Psi\sim r^{-\Delta}
\label{eq:asymptotic_power_ansatz}
\end{equation}
gives
\begin{equation}
\frac{d\Psi}{dr}
=
-\Delta r^{-\Delta-1},
\qquad
\frac{d^2\Psi}{dr^2}
=
\Delta(\Delta+1)r^{-\Delta-2}.
\label{eq:asymptotic_power_derivatives}
\end{equation}

Substitution into Equation~\eqref{eq:asymptotic_Euler_equation} yields
\begin{equation}
\left[
\Delta(\Delta+1)
-
4\Delta
-
\mu^2L^2
\right]
r^{-\Delta}
=
0,
\label{eq:indicial_intermediate}
\end{equation}
and, therefore,
\begin{equation}
\boxed{
\Delta(\Delta-3)
=
\mu^2L^2.
}
\label{eq:AdS_indicial_equation}
\end{equation}

The two asymptotic exponents are
\begin{equation}
\boxed{
\Delta_\pm
=
\frac{3}{2}
\pm
\sqrt{
\frac{9}{4}
+
\mu^2L^2
}.
}
\label{eq:Delta_pm}
\end{equation}

Reality of the exponents requires the usual
Breitenlohner--Freedman condition
\begin{equation}
\mu^2L^2
\geq
-\frac{9}{4}
\label{eq:BF_bound}
\end{equation}
\cite{BreitenlohnerFreedman1982}.
We adopt the standard AdS quantization throughout, selecting the
$\Delta_+$ branch as the normalizable response.  In~the mass range
where alternative quantization is admissible, that possibility is not
considered in this~analysis.

With this choice, the~asymptotic scalar field takes the form
\begin{equation}
\Psi(\ell,\theta)
\sim
A(\omega,\theta)\,r^{-\Delta_-}
+
B(\omega,\theta)\,r^{-\Delta_+},
\qquad
r\rightarrow\infty.
\label{eq:AdS_scalar_falloff}
\end{equation}

The slower falloff is identified with the source coefficient and the
faster falloff with the normalizable response
~\cite{AvisIshamStorey1978,IshibashiWald2004,
Maldacena1998,GubserKlebanovPolyakov1998,Witten1998}.

Using the asymptotic relation in
Equation~\eqref{eq:r_asymptotic_ell}, the~radial falloffs become
\begin{equation}
r^{-\Delta_\pm}
\propto
e^{-\Delta_\pm|\ell|/L}.
\label{eq:AdS_falloff_ell}
\end{equation}

Projecting Equation~\eqref{eq:AdS_scalar_falloff} onto the harmonic basis
already defined in Equation~\eqref{eq:harmonic_expansion_ads} gives
\begin{equation}
R_L(\ell)
=
A_L(\omega)\,r^{-\Delta_-}
+
B_L(\omega)\,r^{-\Delta_+}
+
\cdots ,
\label{eq:channel_asymptotic_falloff}
\end{equation}
where
\begin{align}
A_L(\omega)
&=
\int_0^\pi
d\theta\,
\sin\theta\,
Y_{Lm}^{*}(\theta)
A(\omega,\theta),
\label{eq:AL_projection}
\\
B_L(\omega)
&=
\int_0^\pi
d\theta\,
\sin\theta\,
Y_{Lm}^{*}(\theta)
B(\omega,\theta).
\label{eq:BL_projection}
\end{align}

For an $N$-channel truncation, the~asymptotic data are, therefore, vectors
\begin{equation}
\mathbf A(\omega)
=
\begin{pmatrix}
A_{L_1}\\
A_{L_2}\\
\vdots\\
A_{L_N}
\end{pmatrix},
\qquad
\mathbf B(\omega)
=
\begin{pmatrix}
B_{L_1}\\
B_{L_2}\\
\vdots\\
B_{L_N}
\end{pmatrix}.
\label{eq:AB_channel_vectors}
\end{equation}

For either parity sector, the~throat conditions leave $N$ independent
regular solutions,
\begin{equation}
\mathbf R_{\rm reg}^{(i)}(\ell,\omega),
\qquad
i=1,\ldots,N.
\label{eq:regular_basis_solutions}
\end{equation}

A general regular solution is
\begin{equation}
\mathbf R_{\rm reg}
=
\sum_{i=1}^{N}
c_i
\mathbf R_{\rm reg}^{(i)},
\label{eq:regular_linear_combination}
\end{equation}
with coefficient vector
\begin{equation}
\mathbf c
=
\begin{pmatrix}
c_1\\
c_2\\
\vdots\\
c_N
\end{pmatrix}.
\label{eq:c_vector}
\end{equation}

Propagating each basis solution from the throat to the asymptotic AdS
region gives
\begin{equation}
\mathbf R_{\rm reg}^{(i)}
\sim
\mathbf A^{(i)}(\omega)\,
r^{-\Delta_-}
+
\mathbf B^{(i)}(\omega)\,
r^{-\Delta_+},
\label{eq:basis_asymptotic_expansion}
\end{equation}
where
\begin{equation}
\mathbf A^{(i)}
=
\begin{pmatrix}
A^{(i)}_{L_1}\\
\vdots\\
A^{(i)}_{L_N}
\end{pmatrix},
\qquad
\mathbf B^{(i)}
=
\begin{pmatrix}
B^{(i)}_{L_1}\\
\vdots\\
B^{(i)}_{L_N}
\end{pmatrix}.
\label{eq:basis_AB_vectors}
\end{equation}

Collecting these asymptotic vectors columnwise defines the matching
matrices
\begin{equation}
\boxed{
\mathbf A(\omega)
=
\left(
\mathbf A^{(1)},
\ldots,
\mathbf A^{(N)}
\right),
\qquad
\mathbf B(\omega)
=
\left(
\mathbf B^{(1)},
\ldots,
\mathbf B^{(N)}
\right).
}
\label{eq:matching_matrices_AB}
\end{equation}

The general regular solution, therefore, has asymptotic form
\begin{equation}
\boxed{
\mathbf R_{\rm reg}
\sim
\mathbf A(\omega)\mathbf c\,
r^{-\Delta_-}
+
\mathbf B(\omega)\mathbf c\,
r^{-\Delta_+}.
}
\label{eq:general_matching_asymptotics}
\end{equation}

Because each parity-sector solution on $\ell\geq0$ extends uniquely to
the second exterior by reflection, the~matrices
$\mathbf A_\pm(\omega)$ and $\mathbf B_\pm(\omega)$ constructed in this
way are the parity-resolved matching matrices for the full two-boundary
wormhole~problem.


A normal mode contains no external source at either asymptotic AdS
boundary.  Equation~\eqref{eq:general_matching_asymptotics}, therefore,
requires
\begin{equation}
\mathbf A_\pm(\omega)\mathbf c=0,
\label{eq:source_free_condition}
\end{equation}
where the subscript denotes the even or odd throat-parity sector.
A nontrivial coefficient vector exists only when
$\mathbf A_\pm(\omega)$ loses rank.  The~global normal-mode
quantization condition is consequently
\begin{equation}
\boxed{
\det \mathbf A_\pm
\left(
\omega;r_0,L,a,\epsilon,\mu,m
\right)
=0.
}
\label{eq:global_determinant_condition}
\end{equation}

Each root of Equation~\eqref{eq:global_determinant_condition} determines a
solution regular at the throat and normalizable at both asymptotic AdS
boundaries.  In~a separable system this condition reduces to independent
scalar equations for the individual harmonic sectors.  Here the
allowed frequencies arise collectively from the rank deficiency of the
full channel-matching~matrix.

The dependence of Equation~\eqref{eq:global_determinant_condition} on the
wormhole geometry is completely inherited from the projected matrices
derived in Section~\ref{sec:geometry_scalar}.  The~construction may be
summarized without introducing any additional dynamical input as
\begin{equation}
\boxed{
\left\{
b,\mathcal A,N,\Omega
\right\}
\longrightarrow
\left\{
\mathbf P,\mathbf W,\mathbf C,\mathbf H
\right\}
\longrightarrow
\left\{
\mathbf R_{\rm reg}^{(i)}
\right\}
\longrightarrow
\mathbf A_\pm(\omega)
\longrightarrow
\det\mathbf A_\pm(\omega)=0.
}
\label{eq:geometry_to_determinant}
\end{equation}

To display this construction explicitly, consider the two-channel
sector and channel vector defined in
Equations~\eqref{eq:two_channel_sector_ads} and \eqref{eq:R13_vector}.
Introduce, for~compactness,
\begin{equation}
\delta(\ell)
\equiv
\epsilon\,q[r(\ell)].
\label{eq:delta_definition}
\end{equation}

Then
\begin{equation}
N(\ell,\theta)
=
N_0(\ell)
+
\delta(\ell)P_2(\cos\theta).
\label{eq:N_delta_form}
\end{equation}

The overlap matrix associated with the quadrupolar deformation is the
specialization of Equation~\eqref{eq:P2_overlap_general},
\begin{equation}
M_{LL'}
=
\int_0^\pi
d\theta\,
\sin\theta\,
Y_{L1}(\theta)
P_2(\cos\theta)
Y_{L'1}(\theta),
\label{eq:M13_definition}
\end{equation}
which evaluates for $L,L'\in\{1,3\}$ to
\begin{equation}
\boxed{
\mathbf M
=
\begin{pmatrix}
-\dfrac{1}{5}
&
\dfrac{3\sqrt{14}}{35}
\\[2mm]
\dfrac{3\sqrt{14}}{35}
&
\dfrac{1}{5}
\end{pmatrix}.
}
\label{eq:M13_matrix}
\end{equation}

In particular,
\begin{equation}
M_{13}=M_{31}
=
\frac{3\sqrt{14}}{35},
\label{eq:M13_offdiag}
\end{equation}
so the off-diagonal interaction is fixed by the
$P_2(\cos\theta)$ deformation~itself.

Since $\rho=r$ for the geometry considered here, the~kinetic matrix
\eqref{eq:P_matrix_ads} becomes
\begin{equation}
\boxed{
\mathbf P(\ell)
=
r^2(\ell)
\left[
N_0(\ell)\mathbf I
+
\delta(\ell)\mathbf M
\right].
}
\label{eq:P_two_channel}
\end{equation}

Explicitly,
\begin{equation}
\mathbf P(\ell)
=
r^2(\ell)
\begin{pmatrix}
N_0-\dfrac{\delta}{5}
&
\dfrac{3\sqrt{14}}{35}\delta
\\[2mm]
\dfrac{3\sqrt{14}}{35}\delta
&
N_0+\dfrac{\delta}{5}
\end{pmatrix}.
\label{eq:P_two_channel_explicit}
\end{equation}

The matrices multiplying $\omega^2$, $\omega$, and~the rotational
$\Omega^2$ contribution contain the inverse-lapse overlap
\begin{equation}
J_{LL'}(\ell)
=
\int_0^\pi
d\theta\,
\sin\theta\,
Y_{L1}(\theta)
\frac{1}{
N_0(\ell)+\delta(\ell)P_2(\cos\theta)
}
Y_{L'1}(\theta).
\label{eq:J_definition}
\end{equation}

Equation~\eqref{eq:J_definition} is exact.  For~the controlled
deformation,
\begin{equation}
\frac{1}{
N_0+\delta P_2
}
=
\frac{1}{N_0}
-
\frac{\delta}{N_0^2}P_2
+
\mathcal O(\epsilon^2),
\label{eq:inverse_lapse_expansion}
\end{equation}
and hence
\begin{equation}
\boxed{
\mathbf J(\ell)
=
\frac{1}{N_0(\ell)}
\left[
\mathbf I
-
\frac{\delta(\ell)}{N_0(\ell)}
\mathbf M
\right]
+
\mathcal O(\epsilon^2).
}
\label{eq:J_matrix_expansion}
\end{equation}

Using Equation~\eqref{eq:W_matrix_ads},
\begin{equation}
\boxed{
\mathbf W(\ell)
=
r^2(\ell)\mathbf J(\ell)
=
\frac{r^2(\ell)}{N_0(\ell)}
\left[
\mathbf I
-
\frac{\delta(\ell)}{N_0(\ell)}
\mathbf M
\right]
+
\mathcal O(\epsilon^2).
}
\label{eq:W_two_channel}
\end{equation}

Similarly, because~$m=1$ and
$\Omega=2a/r^3$, Equation~\eqref{eq:C_matrix_ads} gives
\begin{equation}
\boxed{
\mathbf C(\ell)
=
-\frac{4a}{r(\ell)}
\mathbf J(\ell)
=
-\frac{4a}{r(\ell)N_0(\ell)}
\left[
\mathbf I
-
\frac{\delta(\ell)}{N_0(\ell)}
\mathbf M
\right]
+
\mathcal O(\epsilon^2).
}
\label{eq:C_two_channel}
\end{equation}

For the frequency-independent angular contribution, define
\begin{equation}
K_{LL'}
=
\int_0^\pi
d\theta\,
P_2(\cos\theta)
\left[
\sin\theta\,
\partial_\theta Y_{L1}
\partial_\theta Y_{L'1}
+
\frac{
Y_{L1}Y_{L'1}
}{
\sin\theta
}
\right].
\label{eq:K_matrix_definition}
\end{equation}

Direct evaluation yields
\begin{equation}
\boxed{
\mathbf K
=
\begin{pmatrix}
\dfrac{1}{5}
&
\dfrac{12\sqrt{14}}{35}
\\[2mm]
\dfrac{12\sqrt{14}}{35}
&
\dfrac{9}{5}
\end{pmatrix}.
}
\label{eq:K_matrix}
\end{equation}

The undeformed angular operator is represented by
\begin{equation}
\boldsymbol{\Lambda}
=
\begin{pmatrix}
2&0\\
0&12
\end{pmatrix},
\label{eq:Lambda_matrix}
\end{equation}
because $L(L+1)=2$ and $12$ for $L=1$ and $L=3$.

Combining the angular, frame-dragging-squared, and~mass terms in
Equation~\eqref{eq:H_matrix_ads} gives the exact expression
\begin{equation}
\mathbf H(\ell)
=
-N_0(\ell)\boldsymbol{\Lambda}
-
\delta(\ell)\mathbf K
+
\frac{4a^2}{r^4(\ell)}
\mathbf J(\ell)
-
\mu^2r^2(\ell)
\left[
N_0(\ell)\mathbf I
+
\delta(\ell)\mathbf M
\right].
\label{eq:H_two_channel_exact}
\end{equation}

Using Equation~\eqref{eq:J_matrix_expansion},
\begin{equation}
\begin{aligned}
\mathbf H(\ell)
={}&
-N_0\boldsymbol{\Lambda}
-
\delta\mathbf K
\\
&+
\frac{4a^2}{
r^4N_0
}
\left[
\mathbf I
-
\frac{\delta}{N_0}\mathbf M
\right]
\\
&-
\mu^2r^2
\left[
N_0\mathbf I
+
\delta\mathbf M
\right]
+
\mathcal O(\epsilon^2).
\end{aligned}
\label{eq:H_two_channel_expanded}
\end{equation}

Equations~\eqref{eq:P_two_channel},
\eqref{eq:W_two_channel},
\eqref{eq:C_two_channel}, and~\eqref{eq:H_two_channel_exact}, therefore, give the complete
geometry-derived two-channel realization of the operator already
defined in Equation~\eqref{eq:matrix_radial_compact}.  Every coefficient is
fixed by
\begin{equation}
r_0,\qquad
L,\qquad
a,\qquad
\epsilon,\qquad
\mu
\label{eq:two_channel_parameters}
\end{equation}
through $r(\ell)$, $N_0(\ell)$, and~$\delta(\ell)$.

For either throat-parity sector, let
$\mathbf R_\pm^{(1)}$ and $\mathbf R_\pm^{(2)}$ denote the two
independent regular solutions obtained from this operator.  Their
source coefficients define
\begin{equation}
\mathbf A_\pm^{(2)}(\omega)
=
\begin{pmatrix}
A_{1,\pm}^{(1)}(\omega)
&
A_{1,\pm}^{(2)}(\omega)
\\
A_{3,\pm}^{(1)}(\omega)
&
A_{3,\pm}^{(2)}(\omega)
\end{pmatrix}.
\label{eq:A_two_channel_matching}
\end{equation}

The corresponding explicit two-channel quantization condition is
\begin{equation}
\boxed{
\det\mathbf A_\pm^{(2)}(\omega)
=
A_{1,\pm}^{(1)}A_{3,\pm}^{(2)}
-
A_{1,\pm}^{(2)}A_{3,\pm}^{(1)}
=
0.
}
\label{eq:two_channel_quantization}
\end{equation}

The four- and six-channel systems used below follow from the identical
construction with
\begin{equation}
L\in\{1,3,5,7\},
\qquad
L\in\{1,3,5,7,9,11\},
\label{eq:four_six_channel_sets}
\end{equation}
respectively.

A complementary structural check is supplied by the matrix Wronskian.
For two solutions $\mathbf R_1$ and $\mathbf R_2$ of the same coupled
radial problem, define
\begin{equation}
\mathcal W[
\mathbf R_1,\mathbf R_2
]
=
\mathbf R_1^\dagger
\mathbf P
\frac{d\mathbf R_2}{d\ell}
-
\left(
\frac{d\mathbf R_1}{d\ell}
\right)^\dagger
\mathbf P
\mathbf R_2.
\label{eq:matrix_Wronskian}
\end{equation}

Differentiation gives
\begin{equation}
\begin{aligned}
\frac{d\mathcal W}{d\ell}
={}&
\mathbf R_1^\dagger
\frac{d}{d\ell}
\left(
\mathbf P
\frac{d\mathbf R_2}{d\ell}
\right)
\\
&-
\left[
\frac{d}{d\ell}
\left(
\mathbf P
\frac{d\mathbf R_1}{d\ell}
\right)
\right]^\dagger
\mathbf R_2,
\end{aligned}
\label{eq:Wronskian_derivative}
\end{equation}
because the remaining derivative terms cancel identically.
Using Equation~\eqref{eq:projected_matrix_system},
\begin{equation}
\frac{d}{d\ell}
\left(
\mathbf P\mathbf R'
\right)
=
-\mathbf Q\mathbf R,
\label{eq:radial_equation_Q_form}
\end{equation}
so that
\begin{equation}
\frac{d\mathcal W}{d\ell}
=
-\mathbf R_1^\dagger
\mathbf Q
\mathbf R_2
+
\left(
\mathbf Q\mathbf R_1
\right)^\dagger
\mathbf R_2.
\label{eq:Wronskian_Q}
\end{equation}

For the conservative normal-mode problem, the~projected coefficient
operator is Hermitian,
\begin{equation}
\mathbf Q^\dagger=\mathbf Q,
\label{eq:Q_Hermitian}
\end{equation}
and, therefore,
\begin{equation}
\boxed{
\frac{d\mathcal W}{d\ell}=0.
}
\label{eq:Wronskian_conservation}
\end{equation}

The conserved matrix Wronskian is a structural property of the
differential operator and provides a consistency diagnostic for
numerical propagation~\cite{Zettl2005,Teschl2014}.  It should not be
confused with Equation~\eqref{eq:global_determinant_condition}, which is the
global boundary condition that selects the physical normal-mode
frequencies.

The coupled determinant also admits a simple analytic interpretation.
For two weakly interacting spectral branches, write the reduced
matching matrix as
\begin{equation}
\mathbf A_{\rm eff}(\omega)
=
\begin{pmatrix}
A_1(\omega)
&
\epsilon C_{12}(\omega)
\\
\epsilon C_{21}(\omega)
&
A_2(\omega)
\end{pmatrix}.
\label{eq:effective_matching_matrix}
\end{equation}

Its determinant is
\begin{equation}
A_1(\omega)A_2(\omega)
-
\epsilon^2
C_{12}(\omega)C_{21}(\omega)
=
0.
\label{eq:effective_two_channel_det}
\end{equation}

For $\epsilon=0$, the~two channels satisfy
\begin{equation}
A_1(\omega_1^{(0)})=0,
\qquad
A_2(\omega_2^{(0)})=0.
\label{eq:uncoupled_roots}
\end{equation}

Away from degeneracy, consider a mode perturbing from
$\omega_1^{(0)}$,
\begin{equation}
\omega
=
\omega_1^{(0)}
+
\delta\omega,
\qquad
|\delta\omega|\ll1.
\label{eq:weak_frequency_expansion}
\end{equation}

Then
\begin{equation}
A_1(\omega)
\simeq
A_1'
\left(
\omega_1^{(0)}
\right)
\delta\omega,
\label{eq:A1_expansion}
\end{equation}
while
\begin{equation}
A_2(\omega)
\simeq
A_2
\left(
\omega_1^{(0)}
\right).
\label{eq:A2_expansion}
\end{equation}

Equation~\eqref{eq:effective_two_channel_det} consequently gives
\begin{equation}
\boxed{
\delta\omega
=
\epsilon^2
\frac{
C_{12}(\omega_1^{(0)})
C_{21}(\omega_1^{(0)})
}{
A_1'(\omega_1^{(0)})
A_2(\omega_1^{(0)})
}.
}
\label{eq:perturbative_frequency_shift}
\end{equation}

Thus, the leading nondegenerate shift is second order in the
off-diagonal coupling.  As~two uncoupled roots approach one another,
the perturbative denominator becomes small and the interacting
subspace must instead be diagonalized~nonperturbatively.

Near such a pair, the~local spectral structure is represented by
\begin{equation}
\mathbf M_{\rm eff}
=
\begin{pmatrix}
\omega-\omega_1^{(0)}
&
-\epsilon g
\\
-\epsilon g
&
\omega-\omega_2^{(0)}
\end{pmatrix},
\label{eq:avoided_crossing_matrix}
\end{equation}
where $g$ is the effective coupling strength.  Defining the detuning
\begin{equation}
\Delta
=
\omega_1^{(0)}
-
\omega_2^{(0)},
\label{eq:detuning}
\end{equation}
the condition
\begin{equation}
\det\mathbf M_{\rm eff}=0
\label{eq:effective_det_zero}
\end{equation}
gives
\begin{equation}
\left(
\omega-\omega_1^{(0)}
\right)
\left(
\omega-\omega_2^{(0)}
\right)
-
\epsilon^2g^2
=
0,
\label{eq:avoided_crossing_quadratic}
\end{equation}
with exact roots
\begin{equation}
\boxed{
\omega_\pm
=
\frac{
\omega_1^{(0)}
+
\omega_2^{(0)}
}{2}
\pm
\sqrt{
\left(
\frac{\Delta}{2}
\right)^2
+
\epsilon^2g^2
}.
}
\label{eq:avoided_crossing_roots}
\end{equation}

For
\begin{equation}
|\epsilon g|\ll|\Delta|,
\label{eq:weak_coupling_detuning}
\end{equation}
one may expand
\begin{equation}
\sqrt{
\frac{\Delta^2}{4}
+
\epsilon^2g^2
}
=
\frac{|\Delta|}{2}
+
\frac{\epsilon^2g^2}{|\Delta|}
+
\mathcal O(\epsilon^4),
\label{eq:sqrt_weak_expansion}
\end{equation}
which reproduces the second-order perturbative shifts of
Equation~\eqref{eq:perturbative_frequency_shift}.  At~exact resonance,
\begin{equation}
\omega_1^{(0)}
=
\omega_2^{(0)}
=
\omega_0,
\label{eq:exact_resonance}
\end{equation}

Equation~\eqref{eq:avoided_crossing_roots} becomes
\begin{equation}
\omega_\pm
=
\omega_0
\pm
\epsilon g,
\label{eq:resonant_roots}
\end{equation}
so that the minimum spectral gap is
\begin{equation}
\boxed{
\Delta\omega_{\min}
=
2\epsilon g.
}
\label{eq:min_spectral_gap}
\end{equation}

The corresponding eigenvectors may be written as
\begin{equation}
\begin{pmatrix}
u_+\\
u_-
\end{pmatrix}
=
\begin{pmatrix}
\cos\Theta & \sin\Theta\\
-\sin\Theta & \cos\Theta
\end{pmatrix}
\begin{pmatrix}
u_1\\
u_2
\end{pmatrix},
\label{eq:mixing_rotation}
\end{equation}
where
\begin{equation}
\tan(2\Theta)
=
-\frac{2\epsilon g}{\Delta}.
\label{eq:mixing_angle}
\end{equation}
At exact resonance,
\begin{equation}
\Theta=\frac{\pi}{4},
\label{eq:maximal_mixing_angle}
\end{equation}
and, therefore,
\begin{equation}
u_+
=
\frac{1}{\sqrt{2}}
\left(
u_1+u_2
\right),
\qquad
u_-
=
\frac{1}{\sqrt{2}}
\left(
-u_1+u_2
\right).
\label{eq:maximal_mixed_modes}
\end{equation}

\begin{figure}[H]

\includegraphics[width=0.72\textwidth]
{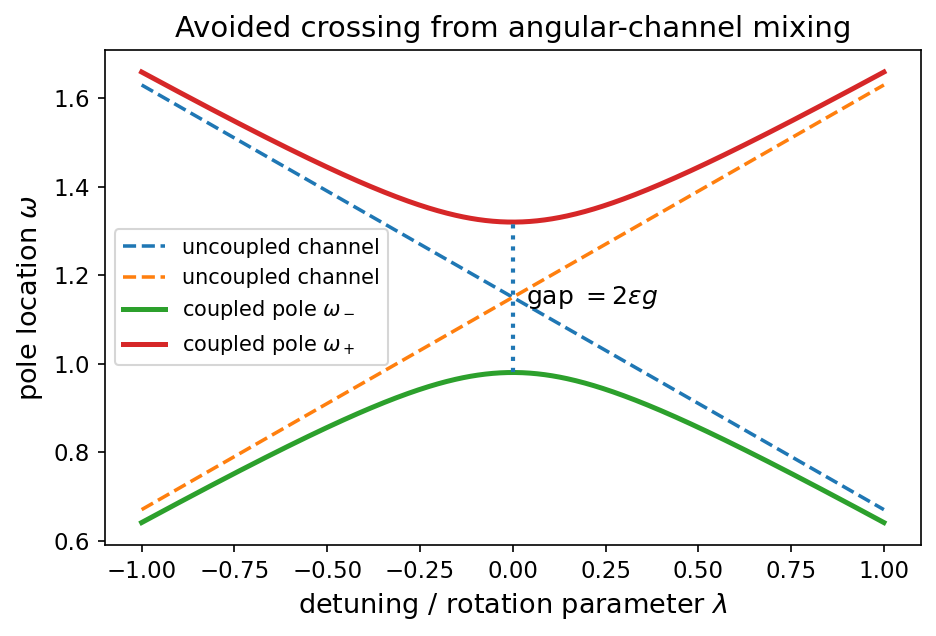}
\caption{Schematic avoided crossing in the two-channel approximation.
Dashed curves denote the uncoupled channel frequencies, while the
solid curves show the coupled normal-mode frequencies obtained from
the determinant condition $\det\mathbf A(\omega)=0$, or~equivalently from the effective $2\times2$ spectral matrix. The~off-diagonal coupling removes the
level crossing, producing the characteristic spectral repulsion with
minimum separation $\Delta\omega=2\epsilon g$ at exact resonance.}
\label{fig:avoided_crossing}
\end{figure}

Thus, the eigenmodes evolve from approximately individual angular
channels far from resonance to collective channel superpositions near
an avoided crossing.  Figure~\ref{fig:avoided_crossing} is an
illustrative realization of this analytic mechanism; it is not a
numerical solution of the full rotating AdS-Teo operator.  The~multi-channel spectrum and its convergence are calculated directly
from the geometry-derived system in Section~\ref{sec:numerical_spectrum}.

The absence of a horizon is essential to the spectral interpretation.
No absorptive condition is imposed at the throat, and~all calculations
are restricted to the ergoregion-free regime
\eqref{eq:ergoregion_global_bound} with reflecting AdS boundary
conditions.  The~resulting computations below yield a discrete
normal-mode spectrum and resolve no nonzero imaginary components
within the numerical precision of the calculations over the tested
parameter range.  This does not imply that an arbitrary quadratic
operator pencil of the form
\eqref{eq:quadratic_operator_pencil} must possess an entirely real
spectrum; rather, it is the numerical spectral behavior obtained for
the geometry, boundary conditions, discretizations, and~parameter
range investigated~here.

\section{Numerical Spectrum and Channel~Convergence}
\label{sec:numerical_spectrum}

We now solve the geometry-derived spectral problem defined by
Equation~\eqref{eq:pencil_eigenproblem}, imposing the odd-parity throat
condition in Equation~\eqref{eq:odd_throat_bc} and the normalizable,
source-free AdS boundary condition associated with the $\Delta_+$ branch
in Equation~\eqref{eq:channel_asymptotic_falloff}.  The~resulting
eigenfrequencies are, therefore, normal modes of the coupled scalar system
under reflecting AdS boundary conditions.  The~angular basis is enlarged
according to Equation~\eqref{eq:channel_hierarchy_ads}, using successively
two, four, and~six retained~channels.

The radial dependence is discretized by a Galerkin expansion on the
half-line computational domain
\begin{equation}
0\leq \ell\leq \ell_{\max}.
\label{eq:numerical_domain}
\end{equation}

In the odd-parity sector used for the production calculations, the~radial basis is the orthonormal sine basis
\begin{equation}
\phi_n(\ell)
=
\sqrt{\frac{2}{\ell_{\max}}}
\sin\!\left(
\frac{n\pi\ell}{\ell_{\max}}
\right),
\qquad
n=1,\ldots,N_r .
\label{eq:radial_galerkin_basis}
\end{equation}

This basis enforces
\begin{equation}
R_L(0)=0,
\qquad
R_L(\ell_{\max})=0,
\label{eq:numerical_radial_bc}
\end{equation}
so the numerical implementation uses a Dirichlet condition at the
finite outer cutoff rather than placing the conformal AdS boundary
itself on the computational grid.  The~finite-cutoff Dirichlet
condition is, therefore, an approximation to the normalizable
source-free $\Delta_+$ condition at $\ell\rightarrow\infty$, and~its
associated finite-domain sensitivity is assessed below through a
combined test in which $\ell_{\max}$ and the radial resolution are
increased together.  Because~the odd-parity solution on $\ell\geq0$ extends
to the second exterior by
$R_L(-\ell)=-R_L(\ell)$, the~condition at
$\ell=+\ell_{\max}$ is mapped to the corresponding Dirichlet
condition at $\ell=-\ell_{\max}$.  Thus, the same finite-cutoff
approximation to the source-free AdS condition is imposed at both
asymptotic ends of the parity-extended numerical~solution.

For the production calculations below we use
\begin{equation}
\ell_{\max}=25,
\qquad
N_r=45,
\label{eq:production_resolution}
\end{equation}
while the convergence studies below examine the radial resolution
independently at fixed $\ell_{\max}$ and then vary $\ell_{\max}$ together
with $N_r$ in a combined finite-domain and radial-resolution test.  The~representative background parameters used for the angular-convergence
\mbox{study are}
\begin{equation}
r_0=1,
\qquad
L_{\rm AdS}=5,
\qquad
a=0.2,
\qquad
\epsilon=0.2,
\qquad
\mu=0,
\qquad
m=1.
\label{eq:numerical_parameters}
\end{equation}

The angular truncations are those already specified in
Equation~\eqref{eq:four_six_channel_sets}.  Except~when the radial resolution,
finite-domain cutoff, or~angular truncation is varied explicitly, the~remaining numerical and background parameters are held~fixed.

Galerkin projection converts the quadratic pencil
Equation~\eqref{eq:quadratic_operator_pencil} into the finite-dimensional
quadratic matrix problem
\begin{equation}
\left(
\mathbf A_0
-\omega\,\mathbf C_0
-\omega^2\mathbf W_0
\right)\mathbf u=0.
\label{eq:discrete_qep}
\end{equation}

We linearize this problem to a generalized eigenvalue problem of
doubled dimension and solve it using the generalized eigensolver
implemented by \texttt{scipy.linalg.eig}.  Only finite generalized
eigenvalues are retained.  For~the low-lying normal-mode analysis,
candidate roots are further restricted to the prescribed positive
frequency window and to
$|\operatorname{Im}\omega|\leq10^{-7}$; this near-reality criterion
is used only for mode selection and continuation and is not used in
the complete complex-spectrum analysis described~below.

For each retained low-lying eigenpair, the~eigenvector is normalized
and its residual is evaluated directly in the original quadratic
problem,
\begin{equation}
\mathcal R(\omega,\mathbf u)
=
\frac{
\left\|
\left(
\mathbf A_0-\omega\mathbf C_0-\omega^2\mathbf W_0
\right)\mathbf u
\right\|
}{
\|\mathbf A_0\mathbf u\|
+
|\omega|\,\|\mathbf C_0\mathbf u\|
+
|\omega|^2\|\mathbf W_0\mathbf u\|
}.
\label{eq:numerical_qep_residual}
\end{equation}

Numerically duplicate roots are removed using a frequency tolerance of
$10^{-7}$, retaining the member of each duplicate pair with the
smaller residual.  During~parameter continuation and convergence
studies, corresponding modes are matched using a cost function that
combines relative frequency separation with the overlap of the
Galerkin eigenvectors; the resulting assignment is obtained by a
minimum-cost bipartite matching procedure.  These residual, duplicate
removal, and~mode-matching tests define the low-lying physical modes
used in the convergence and spectral-flow analyses.  The~complete
complex-spectrum audit below is deliberately less restrictive and
retains every finite generalized eigenvalue in order to test for
additional complex~roots.

For numerical diagnostics, each generalized eigenvalue is substituted
back into the original quadratic matrix pencil and assigned a residual.
Low-lying modes are compared between successive discretizations by
matching both their frequencies and, where angular truncations differ,
their embedded eigenvectors.  This avoids identifying modes solely by
their ordered spectral index when neighboring eigenvalues become
closely spaced.  The~convergence tests below examine the radial basis
size independently at fixed finite domain, the~combined effect of
enlarging the finite computational domain together with the radial
resolution, and~the number of retained angular~channels.

\subsection{Radial-Resolution and Combined Finite-Domain~Convergence}
\label{subsec:radial_domain_convergence}

We first test convergence with respect to the radial discretization.
For fixed $\ell_{\max}=20$, the~number of radial basis functions is
increased through
\begin{equation}
N_r=24,\ 30,\ 36,\ 42,\ 48,\ 54,
\end{equation}
with the $N_r=54$ calculation used as the reference.  The~comparison is
performed for the first 18 matched low-lying modes at both the
representative rotation $a=0.2$ and the near-bound value $a=0.449$.
Table~\ref{tab:radial_convergence} gives the maximum relative frequency
difference with respect to the reference spectrum together with the
maximum quadratic-pencil residual among the matched~modes.

\begin{table}[H]
\centering
\caption{Radial-resolution convergence of the first 18 matched low-lying
normal modes at fixed $\ell_{\max}=20$.  The~$N_r=54$ calculation is
used as the reference at each value of $a$.  The~column
$\max|\Delta\omega/\omega|$ gives the largest relative frequency
difference among the matched modes, while $\mathcal{R}_{\max}$ denotes
the largest quadratic-pencil residual in the same set.
}
\label{tab:radial_convergence}
\begin{tabularx}{\textwidth}{C@{\extracolsep{\fill}}CcCCc}
\toprule
$\boldsymbol{a}$ & $\boldsymbol{N_r}$ & \textbf{Matched Modes} &
$\boldsymbol{\max|\Delta\omega/\omega|}$ &
$\boldsymbol{\mathcal{R}_{\max}}$ \\
\midrule
0.200 & 24 & 18 & $4.287\times10^{-4}$ & $3.648\times10^{-9}$ \\
0.200 & 30 & 18 & $8.282\times10^{-5}$ & $6.905\times10^{-9}$ \\
0.200 & 36 & 18 & $1.260\times10^{-5}$ & $6.934\times10^{-9}$ \\
0.200 & 42 & 18 & $1.027\times10^{-6}$ & $9.263\times10^{-9}$ \\
0.200 & 48 & 18 & $7.675\times10^{-8}$ & $1.581\times10^{-8}$ \\
0.200 & 54 & 18 & $0$                     & $2.378\times10^{-8}$ \\
\midrule
0.449 & 24 & 18 & $5.309\times10^{-4}$ & $6.124\times10^{-9}$ \\
0.449 & 30 & 18 & $1.590\times10^{-4}$ & $7.520\times10^{-9}$ \\
0.449 & 36 & 18 & $1.312\times10^{-5}$ & $9.085\times10^{-9}$ \\
0.449 & 42 & 18 & $6.103\times10^{-7}$ & $1.619\times10^{-8}$ \\
0.449 & 48 & 18 & $4.484\times10^{-8}$ & $1.766\times10^{-8}$ \\
0.449 & 54 & 18 & $0$                     & $2.765\times10^{-8}$ \\
\bottomrule
\end{tabularx}
\end{table}

The frequency differences decrease systematically as the radial basis
is enlarged.  Relative to the $N_r=54$ reference calculation, the~maximum difference among the first 18~matched modes has already fallen
to approximately $10^{-5}$ at $N_r=36$ and below $10^{-6}$ at
$N_r=42$ for both rotation values.  The~residuals remain small
throughout the sequence.  The~same convergence pattern, therefore,
persists from the representative configuration $a=0.2$ to
$a=0.449$, immediately inside the sufficient ergoregion-free~bound.

We next perform a combined finite-domain and radial-resolution test.
Increasing $\ell_{\max}$ at fixed $N_r$ would decrease the radial
resolution of the sine-basis representation, so the radial basis is
enlarged together with the computational domain.  We use
\begin{equation}
(\ell_{\max},N_r)
=
(15,27),\ (20,36),\ (25,45),\ (30,54),
\end{equation}
and compare these calculations with a higher-resolution
$(30,60)$ reference.  This test, therefore, probes the numerical convergence of the
spectrum under simultaneous enlargement of the computational domain
and radial basis; it does not isolate the dependence on $\ell_{\max}$
at fixed $N_r$.  In~addition to the matched low-lying frequencies, we
monitor the representative neighboring-mode gap that will be analyzed
below.

The combined test shows systematic numerical convergence as the
computational domain and radial basis are enlarged together.  Relative
to the $(30,60)$ reference, the~$(15,27)$ calculation differs at the
percent level, whereas the discrepancy decreases to the $10^{-3}$
level for $(20,36)$ and to below $10^{-4}$ for the production choice
$(\ell_{\max},N_r)=(25,45)$.  At~$(30,54)$ the remaining difference is
of order $10^{-5}$.  The~representative neighboring-mode gap also
converges systematically over the same sequence.  Together with the
independent fixed-$\ell_{\max}$ radial-resolution test above, these
results show systematic convergence of the low-lying spectrum as the
finite domain and radial representation are enlarged, while not
treating the combined calculation as an isolated cutoff-convergence
test. The results of this combined convergence test are summarized in Table~\ref{tab:domain_convergence}.

\begin{table}[H]
\centering
\caption{Combined finite-domain and radial-resolution convergence test at
$a=0.2$ and \mbox{$a=0.449$.}  The~radial basis size $N_r$ is enlarged
together with the finite-domain cutoff $\ell_{\max}$, and~the
\mbox{$(\ell_{\max},N_r)=(30,60)$} calculation is used as the reference.
The frequency column gives the maximum relative difference among the
matched low-lying modes, while $\Delta\omega_{\rm gap}$ denotes the
representative neighboring-mode separation monitored in the
convergence test.}
\label{tab:domain_convergence}
\begin{tabularx}{\textwidth}{c@{\extracolsep{\fill}}cccc}
\toprule
$\boldsymbol{a}$ & $\boldsymbol{\ell_{\max}}$ & $\boldsymbol{N_r}$ &
$\boldsymbol{\max|\Delta\omega/\omega|}$ &
$\boldsymbol{\Delta\omega_{\rm gap}}$ \\
\midrule
0.200 & 15 & 27 & $1.864\times10^{-2}$ & $8.093438\times10^{-3}$ \\
0.200 & 20 & 36 & $1.531\times10^{-3}$ & $1.049724\times10^{-2}$ \\
0.200 & 25 & 45 & $9.163\times10^{-5}$ & $1.049971\times10^{-2}$ \\
0.200 & 30 & 54 & $1.283\times10^{-5}$ & $1.050073\times10^{-2}$ \\
0.200 & 30 & 60 & $0$                     & $1.050074\times10^{-2}$ \\
\midrule
0.449 & 15 & 27 & $1.903\times10^{-2}$ & $3.251248\times10^{-3}$ \\
0.449 & 20 & 36 & $1.580\times10^{-3}$ & $2.649144\times10^{-3}$ \\
0.449 & 25 & 45 & $9.638\times10^{-5}$ & $2.643706\times10^{-3}$ \\
0.449 & 30 & 54 & $1.542\times10^{-5}$ & $2.644743\times10^{-3}$ \\
0.449 & 30 & 60 & $0$                     & $2.644737\times10^{-3}$ \\
\bottomrule
\end{tabularx}
\end{table}
\subsection{Angular-Channel~Convergence}
\label{subsec:angular_convergence}

Having established radial-resolution convergence at fixed
$\ell_{\max}$ and systematic convergence under the combined
finite-domain and radial-resolution test, we next examine the
truncation of the coupled angular expansion at the production
resolution \mbox{$(\ell_{\max},N_r)=(25,45)$.}  For~the odd-parity sector with
$m=1$, the~successive channel sets are
\begin{equation}
\{1,3\},
\qquad
\{1,3,5,7\},
\qquad
\{1,3,5,7,9,11\},
\end{equation}
corresponding to two-, four-, and~six-channel calculations,
respectively.

To quantify angular convergence, let $\omega_n^{(N)}$ denote the
matched $n$th eigenfrequency obtained with $N$ retained channels.  We
define
\begin{equation}
\Delta_{2\rightarrow4}^{(n)}
=
\left|
\frac{
\omega_n^{(4)}-\omega_n^{(2)}
}{
\omega_n^{(4)}
}
\right|,
\qquad
\Delta_{4\rightarrow6}^{(n)}
=
\left|
\frac{
\omega_n^{(6)}-\omega_n^{(4)}
}{
\omega_n^{(6)}
}
\right|.
\label{eq:channel_relative_changes}
\end{equation}

Modes are matched between successive truncations using both their
frequency proximity and the overlap of their normalized eigenvectors
after embedding the smaller channel space into the larger one.  We
denote the corresponding embedded overlaps by
$\mathcal{O}_{2\rightarrow4}$ and
$\mathcal{O}_{4\rightarrow6}$.  This provides an additional check that
the frequency comparison follows the same physical mode rather than
merely the same ordered~eigenvalue.

Table~\ref{tab:channel_convergence} shows a clear hierarchy.  Among~the
first ten displayed modes,
\begin{equation}
\max_n\Delta_{2\rightarrow4}^{(n)}
=
4.698\times10^{-3},
\qquad
\max_n\Delta_{4\rightarrow6}^{(n)}
=
2.712\times10^{-5}.
\end{equation}

The corresponding minimum embedded overlaps are
\begin{equation}
\min_n\mathcal{O}_{2\rightarrow4}=0.909133,
\qquad
\min_n\mathcal{O}_{4\rightarrow6}=0.999847.
\end{equation}

Thus, the four-to-six-channel frequency corrections are substantially
smaller than the preceding two-to-four-channel corrections, while the
four-to-six eigenvector overlaps are essentially unity for the
displayed modes.  Fifteen consistent $2\rightarrow4\rightarrow6$
mode chains are identified in the full matching audit.  The~largest
quadratic-pencil residual among the first 18 modes is
$2.114\times10^{-7}$, $1.996\times10^{-7}$, and~$3.062\times10^{-7}$ for the two-, four-, and~six-channel
calculations, respectively.

\begin{table}[H]
\centering
\caption{Angular-channel convergence of the first ten displayed matched
low-lying normal modes for the odd-parity sector at
$(\ell_{\max},N_r)=(25,45)$, with~$a=0.2$, $\epsilon=0.2$, and~$m=1$.
The relative changes are defined in
Equation~\eqref{eq:channel_relative_changes}.  The~last two columns give
the normalized embedded eigenvector overlaps used in matching the
successive truncations.
}
\label{tab:channel_convergence}


\begin{tabularx}{\textwidth}{c@{\extracolsep{\fill}}ccccccc}
\toprule
\textbf{Mode} &
$\boldsymbol{\omega^{(2)}}$ &
$\boldsymbol{\omega^{(4)}}$ &
$\boldsymbol{\omega^{(6)}}$ &
$\boldsymbol{\Delta_{2\rightarrow4}}$ &
$\boldsymbol{\Delta_{4\rightarrow6}}$ &
$\boldsymbol{\mathcal{O}_{2\rightarrow4}}$ &
$\boldsymbol{\mathcal{O}_{4\rightarrow6}}$
\\
\midrule
1  & 0.788442700 & 0.788441157 & 0.788441156 &
$1.957\times10^{-6}$ & $1.187\times10^{-9}$ & 1.000000 & 1.000000 \\
2  & 1.174017041 & 1.173985034 & 1.173985033 &
$2.726\times10^{-5}$ & $2.213\times10^{-10}$ & 0.999984 & 1.000000 \\
3  & 1.194674268 & 1.194262885 & 1.194262882 &
$3.445\times10^{-4}$ & $2.719\times10^{-9}$ & 0.999564 & 1.000000 \\
4  & 1.557445161 & 1.557397493 & 1.557397490 &
$3.061\times10^{-5}$ & $1.811\times10^{-9}$ & 0.999967 & 1.000000 \\
5  & 1.590184699 & 1.584743972 & 1.584743631 &
$3.433\times10^{-3}$ & $2.153\times10^{-7}$ & 0.915347 & 1.000000 \\
6  & 1.939087729 & 1.939013537 & 1.939013525 &
$3.826\times10^{-5}$ & $6.532\times10^{-9}$ & 0.999918 & 1.000000 \\
7  & 1.983477189 & 1.974203197 & 1.974190839 &
$4.698\times10^{-3}$ & $6.260\times10^{-6}$ & 0.909133 & 0.999994 \\
8  & 2.319432364 & 2.319118726 & 2.319118529 &
$1.352\times10^{-4}$ & $8.496\times10^{-8}$ & 0.999475 & 1.000000 \\
9  & 2.373887290 & 2.362998365 & 2.362934289 &
$4.608\times10^{-3}$ & $2.712\times10^{-5}$ & 0.937291 & 0.999847 \\
10 & 2.699202103 & 2.698322012 & 2.698321419 &
$3.262\times10^{-4}$ & $2.198\times10^{-7}$ & 0.998534 & 1.000000 \\
\bottomrule
\end{tabularx}

\end{table}

\begin{figure}[H]

\includegraphics[width=0.9\textwidth]
{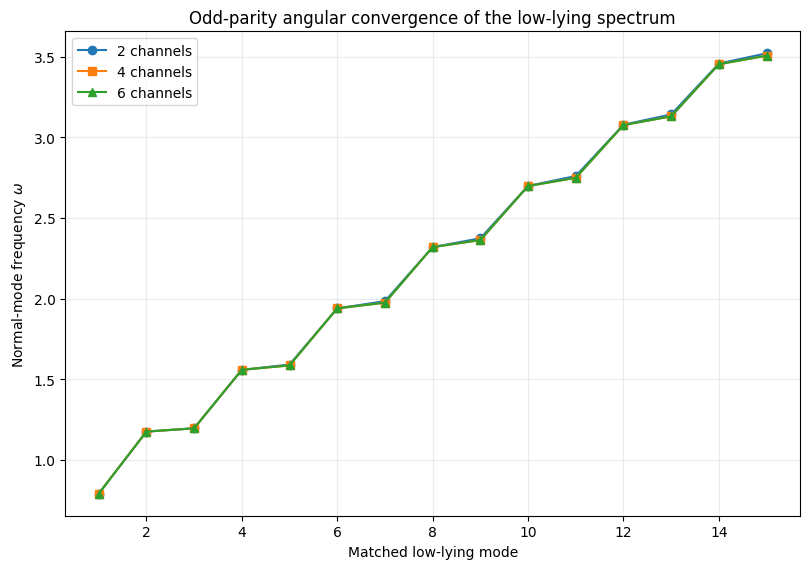}
\caption{Comparison of matched normal-mode frequencies obtained with the
two-, four-, and~six-channel truncations at
$(\ell_{\max},N_r)=(25,45)$.  The~different marker shapes and sizes
are used to keep the nearly coincident datasets visible: circles,
squares, and~triangles denote the two-, four-, and~six-channel
results, respectively.  Connecting lines are included only to guide
the eye.  The~close agreement, particularly between the four- and
six-channel results, demonstrates numerical convergence of the
low-lying spectrum under enlargement of the coupled angular basis.
}
\label{fig:channel_convergence_spectrum}
\end{figure}

The two-channel approximation already captures the overall distribution
of the low-lying spectrum, while the four- and six-channel results are
nearly indistinguishable in Figure~\ref{fig:channel_convergence_spectrum}.  Since the matrices being enlarged are the geometry-derived angular projections defined in
Section~\ref{sec:geometry_scalar}, this convergence test probes the
coupled AdS-Teo operator itself rather than a separate phenomenological
truncation. The corresponding low-lying spectra for the two-, four-, and six-channel truncations are shown in Figure~\ref{fig:channel_convergence_spectrum}.

\begin{figure}[H]

\includegraphics[width=0.9\textwidth]
{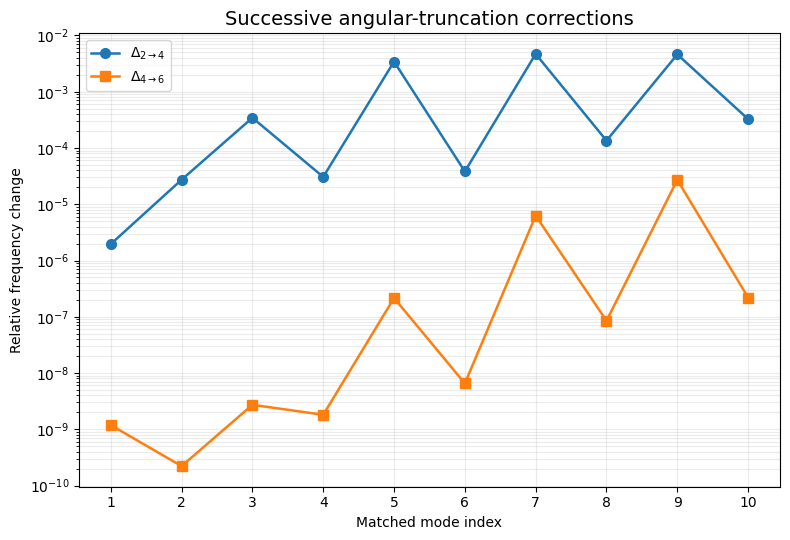}
\caption{Relative changes in the matched normal-mode frequencies under
successive enlargements of the angular basis at
$(\ell_{\max},N_r)=(25,45)$.  Circles denote
$\Delta_{2\rightarrow4}$ and squares denote
$\Delta_{4\rightarrow6}$ from
Equation~\eqref{eq:channel_relative_changes}.  Both are displayed on the
same logarithmic scale, making the substantially smaller
four-to-six-channel corrections directly visible.
}
\label{fig:channel_convergence_relative}
\end{figure}

The corresponding relative changes
$\Delta_{2\rightarrow4}$ and $\Delta_{4\rightarrow6}$ are shown in
Figure~\ref{fig:channel_convergence_relative}.

Taken together, the~fixed-domain radial-resolution test, the~combined
finite-domain and radial-resolution test, and~the angular-truncation
test show systematic numerical convergence of the low-lying
normal-mode spectrum.  We, therefore, adopt the six-channel calculation
with
\begin{equation}
N=6,
\qquad
L\in\{1,3,5,7,9,11\},
\qquad
(\ell_{\max},N_r)=(25,45).
\label{eq:production_discretization}
\end{equation}
as the production discretization for the spectral calculations that
follow.  The~distinction between this numerical convergence and the
question of dynamical mode stability will be addressed separately
below by examining the complete complex \mbox{generalized~eigenspectrum.}

\subsection{Complete Complex-Spectrum~Analysis}
\label{subsec:complex_spectrum_audit}

The generalized eigenvalue problem obtained by linearizing the
quadratic pencil returns both positive- and negative-frequency roots
and, in~principle, may also return genuinely complex frequencies.
With the time dependence
\begin{equation}
\Phi\propto e^{-i\omega t},
\end{equation}
a root with
\begin{equation}
\operatorname{Im}\omega>0
\end{equation}
would correspond to exponential growth in the probe scalar sector.
It is, therefore, necessary to inspect the complete finite
eigenspectrum rather than infer spectral reality from a plot of
selected positive-frequency~modes.

In the production analysis used to display nearly real normal modes,
we employed the conservative numerical selection criterion
\begin{equation}
|\operatorname{Im}\omega|\leq 10^{-7}.
\label{eq:imaginary_selection_tolerance}
\end{equation}

This criterion is useful for selecting modes for plotting and
continuation, but~it cannot by itself establish that additional
complex roots are absent.  We, therefore, performed an independent
audit in which no imaginary-part filter was applied before the
finite generalized eigenspectrum was~examined.

At the production resolution
\begin{equation}
(\ell_{\max},N_r)=(25,45),
\end{equation}
the full finite eigenspectrum was computed at 91 values of the
rotation parameter,
\begin{equation}
a=0,\ 0.005,\ 0.010,\ldots,0.445,\ 0.449,
\label{eq:complex_spectrum_rotation_grid}
\end{equation}
with $\epsilon=0.2$ held fixed.  Each calculation returned 540 finite
generalized eigenvalues, so that the audit examined a total of
\begin{equation}
N_{\rm eig}=91\times540=49{,}140
\label{eq:complex_spectrum_total}
\end{equation}
finite roots.  No nonzero imaginary component was resolved anywhere
in this sweep at the numerical precision of the calculation.  In~particular, the~numerical audit returned
\begin{equation}
N\!\left(|\operatorname{Im}\omega|>0\right)=0,
\qquad
N\!\left(\operatorname{Im}\omega>0\right)=0,
\label{eq:no_complex_roots}
\end{equation}
and
\begin{equation}
\max |\operatorname{Im}\omega|=0
\label{eq:max_imaginary_part}
\end{equation}
to the numerical precision reported by the eigensolver.  These zeros
should, therefore, be understood as numerical results rather than exact
analytical statements.  The~same numerical conclusion is obtained
when the diagnostic thresholds are set at
$10^{-14}$, $10^{-12}$, $10^{-10}$, and~$10^{-8}$.

To test whether this numerical conclusion depends on the radial
discretization or the finite computational cutoff, we repeated the
unfiltered complex-spectrum calculation at the representative rotation
values
\begin{equation}
a=0,\quad 0.2,\quad 0.3945,\quad 0.449
\end{equation}
for the four discretizations
\begin{equation}
(\ell_{\max},N_r)
=
(20,36),\quad
(25,45),\quad
(30,54),\quad
(30,60).
\label{eq:complex_audit_resolutions}
\end{equation}

The corresponding finite generalized eigenspectra contain,
respectively, 432, 540, 648, and~720 roots at each value of $a$.
No nonzero imaginary component was resolved within numerical precision
in any of these 16 calculations.  Thus, the numerical absence of
$\operatorname{Im}\omega>0$ modes is unchanged under simultaneous
variation of radial resolution and finite-domain cutoff, including
near the upper end of the ergoregion-free parameter~interval.

The residual of a computed eigenpair $(\omega,\mathbf{u})$ is
evaluated directly in the original quadratic pencil rather than in
the doubled linearized problem.  In~particular, for~\begin{equation}
\mathbf{P}(\omega)
=
\mathbf{M}_2\omega^2
+
\mathbf{M}_1\omega
+
\mathbf{M}_0,
\end{equation}
we monitor the normalized residual associated with
$\mathbf{P}(\omega)\mathbf{u}$.  The~convergence statements for the
physical low-lying spectrum are based on the matched modes examined
in the radial, combined finite-domain and radial-resolution, and~angular-convergence tests above.  The~complete complex-spectrum audit,
by contrast, deliberately retains every finite generalized eigenvalue
in order to test for additional complex roots.  Consequently, the~largest residual over the entire unfiltered eigenspectrum should not
be interpreted as the accuracy of the converged low-lying~modes.

For example, in~the radial-convergence audit the maximum residuals
among the first 18 matched low-lying modes at the highest tested
resolution are
\begin{equation}
2.378\times10^{-8}
\qquad (a=0.2),
\end{equation}
and
\begin{equation}
2.765\times10^{-8}
\qquad (a=0.449),
\end{equation}
while the angular-convergence audit at the production discretization
gives maximum residuals of order $10^{-7}$ among the first 18 modes.
These residual tests, together with the systematic frequency
convergence demonstrated above, provide the numerical criterion for
identifying the low-lying modes used in the subsequent spectral
analysis.

We, therefore, find no numerical evidence for an exponentially growing
normal mode in the probe scalar sector over the parameter range and
discretizations examined here.  More precisely, no nonzero imaginary
component is resolved in the complete finite-dimensional generalized
eigenspectrum at the numerical precision of the calculations, and~no
convergent mode with $\operatorname{Im}\omega>0$ is found under the
tested changes of radial resolution and finite-domain cutoff.  This
statement concerns only the probe scalar field governed by the
linearized Klein--Gordon problem studied in this work.  It does not
establish gravitational stability of the wormhole geometry or
stability of the matter sector required to support \mbox{the background.}

\subsection{Rotation-Dependent Spectral~Evolution}
\label{subsec:spectral_flow}

We next examine how the six-channel normal-mode spectrum evolves with
the rotation parameter $a$.  Throughout this analysis the deformation
parameter is fixed at $\epsilon=0.2$, while $a$ is varied over the
ergoregion-free interval established in
Section~\ref{sec:geometry_scalar}.  Unless~otherwise stated, the~production discretization defined in Equation~\eqref{eq:production_discretization} is~used.

\begin{figure}[H]
\includegraphics[width=1\textwidth]
{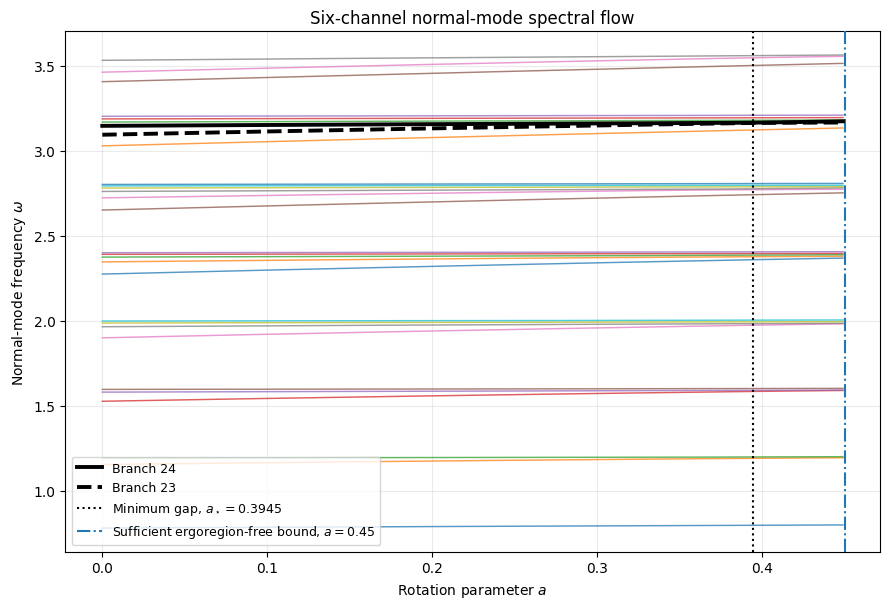}
\caption{Rotation-dependent six-channel normal-mode spectrum for
$\epsilon=0.2$ using the production discretization
$(\ell_{\max},N_r)=(25,45)$.  The~calculation uses the odd-parity
channel set $L\in\{1,3,5,7,9,11\}$.  Branches 24 and 23 are highlighted
to identify the closest finite-gap feature examined below. The remaining colored curves represent the other continuously tracked positive-frequency branches of the six-channel spectrum.}  
The~vertical dashed line at $a=0.45$ indicates the sufficient
ergoregion-free bound derived for the adopted parameters; the
numerical continuation terminates at $a=0.449$.
\label{fig:spectral_flow}
\end{figure}

Figure~\ref{fig:spectral_flow} shows the resulting low-lying
positive-frequency branches.  The~spectrum evolves continuously with
$a$, but~the branches do not all respond uniformly to increasing
rotation.  Several neighboring modes approach one another closely
before separating again.  The~most pronounced feature in the
displayed spectral window occurs between Branches 24 and~23.

To resolve this feature quantitatively, we define the branch
separation
\begin{equation}
\Delta\omega(a)
=
\left|
\omega_{24}(a)-\omega_{23}(a)
\right|.
\label{eq:branch_gap}
\end{equation}

In the dedicated convergence audit at the production
resolution $(\ell_{\max},N_r)=(25,45)$, the local quadratic refinement
through the sampled minimum gives
\begin{equation}
a_\star \simeq 0.3943,
\qquad
\Delta\omega_{\min}
\simeq
1.689\times10^{-3}.
\label{eq:production_gap}
\end{equation}

The nonzero minimum followed by renewed separation is characteristic
of finite level repulsion in the truncated coupled~spectrum.

Because the minimum gap is substantially smaller than the individual
mode frequencies, we performed a dedicated convergence audit of both
its location and magnitude.  The~same Branch 24/23 pair was tracked
at successively larger radial domains and radial resolutions.  The~results are summarized in Table~\ref{tab:gap_convergence}.

\vspace{-6pt}
\begin{table}[H]
\centering
\caption{Convergence of the Branch 24/23 minimum spectral gap under changes
of the finite radial domain $\ell_{\max}$ and radial resolution
$N_r$.  The~relative errors are evaluated with respect to the
highest-resolution calculation $(\ell_{\max},N_r)=(30,60)$.}
\label{tab:gap_convergence}
\begin{tabularx}{\textwidth}{c@{\extracolsep{\fill}}cccc}
\toprule
$\boldsymbol{(\ell_{\max},N_r)}$ &
$\boldsymbol{a_\star}$ &
$\boldsymbol{\Delta\omega_{\min}}$ &
$\boldsymbol{\delta a_\star/a_\star}$ &
$\boldsymbol{\delta(\Delta\omega_{\min})/\Delta\omega_{\min}}$
\\
\midrule
$(20,36)$ &
0.384588 &
$1.934814\times10^{-3}$ &
$1.052\times10^{-2}$ &
$1.567\times10^{-1}$
\\
$(25,45)$ &
0.394326 &
$1.689056\times10^{-3}$ &
$7.776\times10^{-4}$ &
$9.787\times10^{-3}$
\\
$(30,54)$ &
0.394899 &
$1.673475\times10^{-3}$ &
$2.038\times10^{-4}$ &
$4.727\times10^{-4}$
\\
$(30,60)$ &
0.395103 &
$1.672685\times10^{-3}$ &
0 &
0
\\
\bottomrule
\end{tabularx}
\end{table}

The convergence is systematic.  Relative to the
$(30,60)$ reference calculation, the~production discretization locates
the minimum to within approximately $7.8\times10^{-4}$ in relative
$a_\star$ and the minimum gap to within approximately
$9.8\times10^{-3}$.  Increasing the resolution to $(30,54)$ reduces
these discrepancies to approximately $2.0\times10^{-4}$ and
$4.7\times10^{-4}$, respectively.  At~the highest tested resolution,
\begin{equation}
a_\star \simeq 0.3951,
\qquad
\Delta\omega_{\min}
\simeq
1.673\times10^{-3}.
\label{eq:reference_gap}
\end{equation}

Thus, both the existence and magnitude of the finite minimum, as~well
as its location in the rotation parameter, persist under simultaneous
enlargement of the radial domain and radial~resolution.

\begin{figure}[H]

\includegraphics[width=0.78\textwidth]
{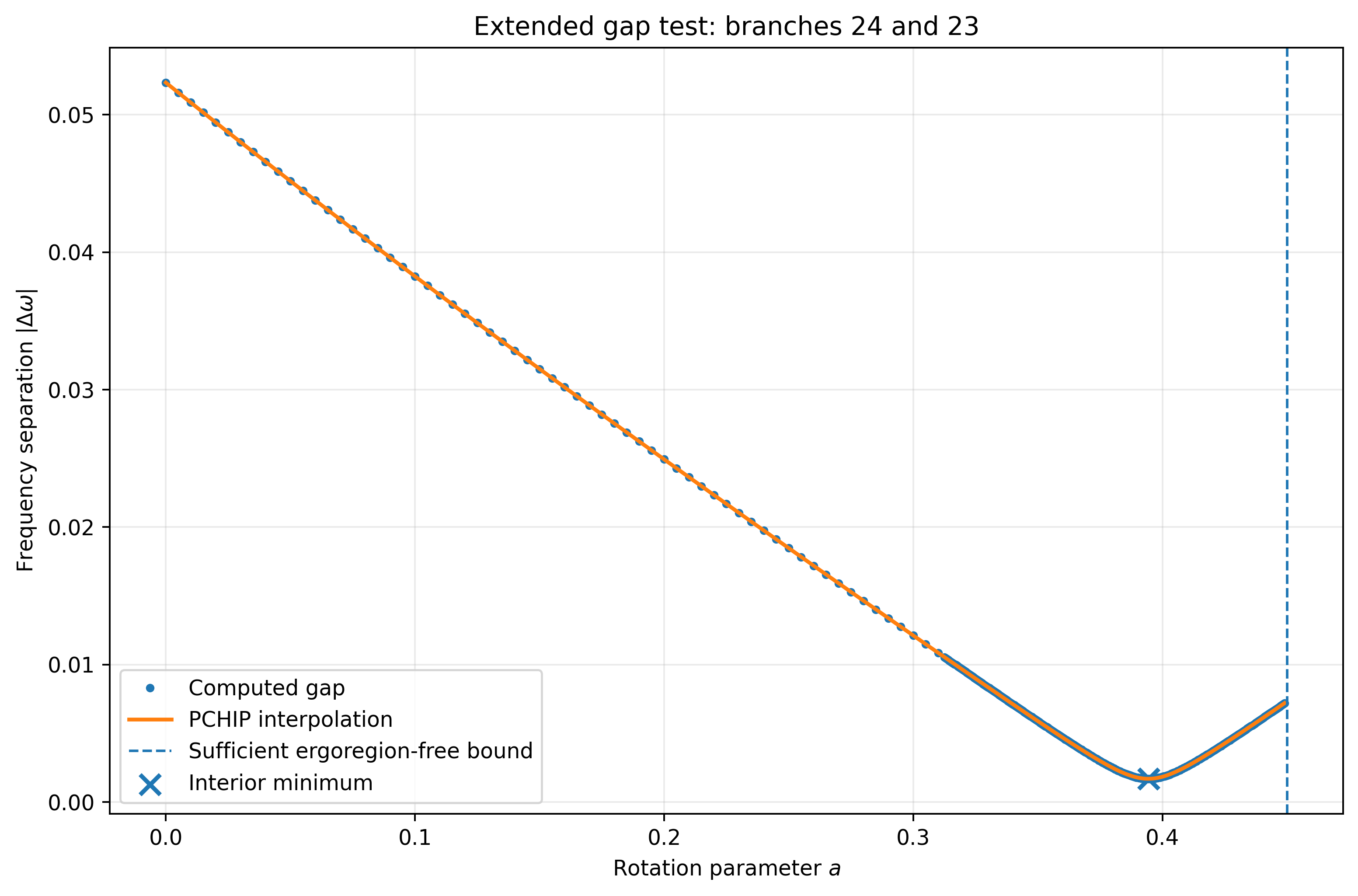}
\caption{Frequency separation $\Delta\omega=|\omega_{24}-\omega_{23}|$ between Branches 24 and 23 in the neighborhood of their closest approach.  The~production calculation gives
$a_\star\simeq0.3943$ and
$\Delta\omega_{\min}\simeq1.689\times10^{-3}$.
The dedicated resolution audit in
Table~\ref{tab:gap_convergence} shows convergence toward
$a_\star\simeq0.3951$ and
$\Delta\omega_{\min}\simeq1.673\times10^{-3}$.
}
\label{fig:spectral_gap}
\end{figure}

The corresponding Branch 24/23 frequency separation
near the closest approach is shown in
Figure~\ref{fig:spectral_gap}.

The gap alone does not determine how the angular content of the
eigenmodes changes through the interaction region.  We, therefore,
inspect the normalized channel weights.  For~a normalized
six-channel eigenvector
\begin{equation}
\mathbf{u}^{(n)}
=
\left(
\mathbf{u}^{(n)}_{1},
\mathbf{u}^{(n)}_{3},
\mathbf{u}^{(n)}_{5},
\mathbf{u}^{(n)}_{7},
\mathbf{u}^{(n)}_{9},
\mathbf{u}^{(n)}_{11}
\right),
\end{equation}
we define
\begin{equation}
W_L^{(n)}
=
\frac{
\left\|
\mathbf{u}^{(n)}_L
\right\|^2
}{
\displaystyle
\sum_{L'}
\left\|
\mathbf{u}^{(n)}_{L'}
\right\|^2
},
\qquad
\sum_L W_L^{(n)}=1.
\label{eq:channel_weights}
\end{equation}

Here $\mathbf{u}^{(n)}_L$ denotes the radial coefficient vector
associated with angular channel $L$.

\begin{figure}[H]

\includegraphics[width=0.82\textwidth]
{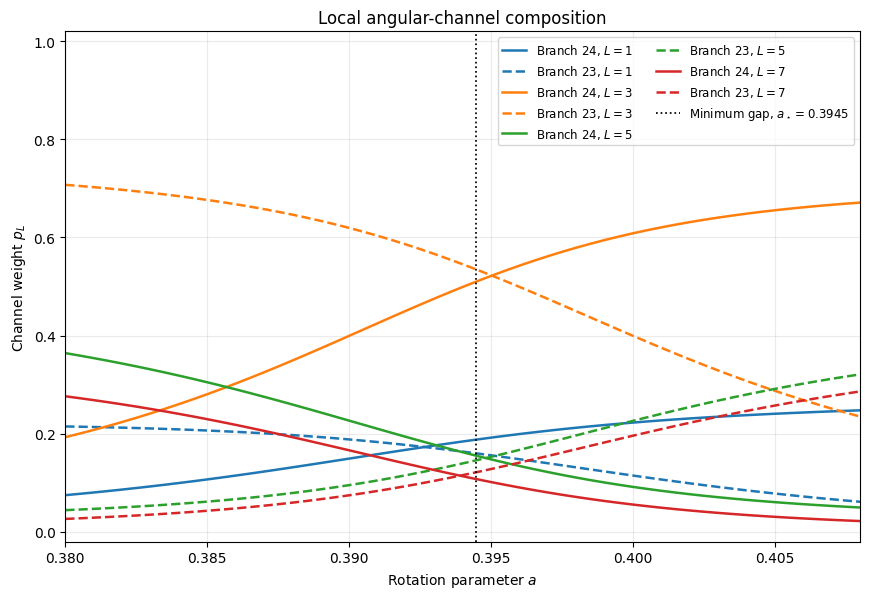}
\caption{Normalized angular-channel weights for Branches 24 and 23 near their
closest approach.  The~dominant $L=1,3,5,7$ contributions are shown;
the remaining $L=9,11$ components are smaller in the displayed
window.  The~vertical marker indicates the minimum obtained from the
production-resolution dataset, $a_\star\simeq0.3945$; the dedicated convergence audit gives the corresponding converged location
$a_\star\simeq0.3951$.  The~redistribution of weight
through the minimum-gap region demonstrates that the spectral feature
is accompanied by a reorganization of the coupled angular content
rather than a literal exchange between two isolated channels.}
\label{fig:branch_weights}
\end{figure}

Figure~\ref{fig:branch_weights} displays the dominant channel weights
for Branches 24 and 23 across the neighborhood of the minimum gap.
The angular composition changes appreciably as the branches traverse
the interaction region.  Importantly, the~numerical eigenvectors do
not exhibit a simple one-for-one exchange between two isolated
angular channels.  Instead, several components participate in the
reorganization.

The combined frequency and eigenvector information, therefore, supports
a conservative interpretation of this feature as a
\emph{collective multichannel spectral reorganization}.  The~finite
minimum of $\Delta\omega$ establishes level repulsion between the two
tracked spectral branches, while the simultaneous redistribution of
several angular components shows that the underlying eigenstates are
genuinely multichannel.  We consequently avoid interpreting the
numerical result as an exact two-state or two-channel avoided~crossing.

A reduced two-level description can nevertheless provide a useful
local language for the observed spectral geometry.  If~two effective
uncoupled levels $\omega_1(a)$ and $\omega_2(a)$ interact through an
effective coupling $g(a)$, one may write
\begin{equation}
H_{\rm eff}(a)
=
\begin{pmatrix}
\omega_1(a) & g(a)\\
g(a) & \omega_2(a)
\end{pmatrix},
\label{eq:effective_two_level}
\end{equation}
whose eigenvalues are
\begin{equation}
\omega_{\pm}(a)
=
\frac{\omega_1+\omega_2}{2}
\pm
\sqrt{
\left(
\frac{\omega_1-\omega_2}{2}
\right)^2
+
g^2
}.
\label{eq:effective_two_level_eigenvalues}
\end{equation}

Near an uncoupled degeneracy, this model produces the familiar
minimum separation
\begin{equation}
\Delta\omega_{\min}\simeq 2|g|.
\end{equation}

For the converged Branch 24/23 gap, this would correspond
schematically to the effective scale
\begin{equation}
|g_{\rm eff}|
\sim
\frac{\Delta\omega_{\min}}{2}
\simeq
8.365\times10^{-4}.
\label{eq:effective_gap_scale}
\end{equation}

This estimate, obtained from the converged minimum gap, should be
understood only as a qualitative local parameterization of the finite
gap.  The~eigenvectors displayed in
Figure~\ref{fig:branch_weights} are obtained from the production-resolution
six-channel calculation and contain several participating angular
channels, so Equation~\eqref{eq:effective_two_level} is not a reduction of
the full six-channel bulk~operator.

The spectral-flow calculation, therefore, establishes three related
features.  First, the~low-lying six-channel spectrum evolves smoothly
throughout the ergoregion-free rotation interval considered here.
Second, neighboring branches can undergo narrow but numerically
resolved finite-gap interactions.  Third, the~associated eigenvectors
undergo multichannel redistribution through these interaction
regions.  The~dedicated convergence test demonstrates that the
Branch 24/23 minimum persists under simultaneous enlargement of the
finite radial domain and radial resolution, with~both its location
and magnitude showing systematic numerical~convergence.

\subsection{Numerical Interpretation and~Scope}
\label{subsec:numerical_interpretation}

The numerical results presented above provide several complementary
checks on the coupled spectral formulation.  The~fixed-domain
radial-resolution study establishes convergence with increasing
$N_r$, while the combined finite-domain and radial-resolution study
shows systematic numerical convergence as $\ell_{\max}$ and $N_r$ are
enlarged together. The~two-, four-, and~six-channel comparison shows
systematic convergence under enlargement of the angular basis.  The~complete unfiltered generalized-eigenspectrum audit resolves no
complex roots within numerical precision over the tested rotation
interval and discretizations.  Finally,
the rotation-dependent calculation identifies convergent finite-gap
spectral interactions accompanied by redistribution of the
angular-channel~content.

These observations should be interpreted within the scope of this
calculation.  In~particular, the~absence of resolved complex
frequencies within numerical precision and the absence of convergent
modes with $\operatorname{Im}\omega>0$ do not constitute a proof of
full gravitational stability.  Such a statement would require
perturbing the metric and the matter sector supporting the wormhole
and analyzing the resulting coupled dynamical system.  Our result is
instead that, within~the linear probe Klein--Gordon problem and over
the parameter range investigated here, no exponentially growing
scalar normal mode is~found.

Likewise, the~finite-gap behavior of Branches 24 and 23 should not be
identified with a literal two-state avoided crossing.  The~six-channel eigenvectors show that the mode composition is
multicomponent, and~the reduced two-level model introduced above is
used only to characterize the local spectral geometry.  The~geometry-derived coupled operator itself remains the basis of all
numerical~results.

The principal numerical conclusions can, therefore, be summarized as
follows:
\begin{enumerate}
\item
the low-lying spectrum converges systematically with radial
resolution at fixed finite domain, shows systematic numerical
convergence under simultaneous enlargement of the finite domain and
radial basis, and~converges under enlargement of the angular-channel
truncation;

\item
the complete finite generalized eigenspectrum remains real at the
numerical precision of the calculations throughout the tested
ergoregion-free parameter range;

\item
the six-channel spectrum exhibits resolved finite-gap interactions
between neighboring branches;

\item
the most closely examined Branch 24/23 feature converges toward
\begin{equation}
a_\star \simeq 0.3951,
\qquad
\Delta\omega_{\min}
\simeq
1.673\times10^{-3},
\end{equation}
at the highest tested resolution $(\ell_{\max},N_r)=(30,60)$; and

\item
the channel-weight evolution through this feature demonstrates
collective multichannel spectral reorganization rather than a
simple exchange of two isolated angular components.
\end{enumerate}

These results provide the numerical foundation for the boundary
response analysis developed in the following~section.

\section{Two-Boundary Response and Quantum~Observables}
\label{sec:boundary_quantum}

The asymptotic coefficients introduced in
Equation~\eqref{eq:matching_matrices_AB} provide the formal ingredients for
constructing the linear response of the two asymptotic AdS
boundaries.  In~this section we derive the corresponding
matrix-valued source-response structure and its relation to the
source-free normal-mode condition.  We do not numerically extract the
full response matrix from the asymptotic coefficients of the
six-channel bulk solutions obtained in
Section~\ref{sec:numerical_spectrum}; the response plots presented
below instead illustrate the formal construction using the reduced
two-channel model introduced earlier.  Because~the wormhole possesses
distinct left and right conformal boundaries, the~corresponding
channel-space source and response vectors are assembled as
\begin{equation}
\mathbb A
=
\begin{pmatrix}
\mathbf A_L\\
\mathbf A_R
\end{pmatrix},
\qquad
\mathbb B
=
\begin{pmatrix}
\mathbf B_L\\
\mathbf B_R
\end{pmatrix}.
\label{eq:global_boundary_vectors}
\end{equation}

The complete linear response is, therefore,
\begin{equation}
\boxed{
\mathbb B
=
\mathbb G(\omega)\,
\mathbb A,
}
\label{eq:global_two_boundary_response}
\end{equation}
where $\mathbb G(\omega)$ acts simultaneously on boundary
(left/right) and angular-channel indices.  In~block form,
\begin{equation}
\boxed{
\mathbb G(\omega)
=
\begin{pmatrix}
\mathbf G_{LL}(\omega)
&
\mathbf G_{LR}(\omega)
\\
\mathbf G_{RL}(\omega)
&
\mathbf G_{RR}(\omega)
\end{pmatrix}.
}
\label{eq:two_boundary_blocks}
\end{equation}

The diagonal blocks describe same-boundary response, while
$\mathbf G_{LR}$ and $\mathbf G_{RL}$ encode propagation through the
wormhole between the two asymptotic regions.  Each block is itself a
matrix in the retained angular-channel basis, so off-diagonal entries
within a block measure angular-channel mixing generated by the
non-separable~geometry.

Reflection symmetry permits the same response problem to be resolved
into the even- and odd-parity sectors introduced in
Equation~\eqref{eq:radial_parity}.  Using the matching matrices already
constructed in Equation~\eqref{eq:matching_matrices_AB}, the~parity-resolved
source and response vectors satisfy
\begin{equation}
\mathbf J_\pm(\omega)
=
\mathbf A_\pm(\omega)\mathbf c,
\qquad
\mathbf O_\pm(\omega)
=
\mathbf B_\pm(\omega)\mathbf c.
\label{eq:parity_source_response}
\end{equation}

Away from the normal-mode frequencies,
\begin{equation}
\mathbf c
=
\mathbf A_\pm^{-1}(\omega)\mathbf J_\pm,
\label{eq:c_from_source}
\end{equation}
and, therefore,
\begin{equation}
\boxed{
\mathbf G_\pm(\omega)
\propto
\mathbf B_\pm(\omega)
\mathbf A_\pm^{-1}(\omega).
}
\label{eq:parity_green_function}
\end{equation}

This is the coupled-channel analogue of the standard AdS
source-response prescription~\cite{Maldacena1998,Witten1998,Aharony2000,Skenderis2002,
SonStarinets2002,KaminskiOperatorMixing2010}.

The poles of Equation~\eqref{eq:parity_green_function} are fixed by the
source-free condition already derived in
Equation~\eqref{eq:global_determinant_condition}.  Thus, the normal-mode
frequencies $\omega_n^{(\pm)}$ are precisely the frequencies at which
the parity-resolved response becomes singular, up~to exceptional
pole-zero cancellations.  Near~an isolated simple pole,
\begin{equation}
\mathbf G_\pm(\omega)
=
\frac{
\mathbf R_n^{(\pm)}
}{
\omega-\omega_n^{(\pm)}
}
+
\mathbf G_{\rm reg}(\omega),
\label{eq:green_pole_expansion}
\end{equation}
where $\mathbf R_n^{(\pm)}$ is the residue matrix.  The~pole location
gives the normal-mode frequency, while the residue records how the
collective mode couples the retained source and \mbox{response~channels.}

For real background coefficients, complex conjugation relates the
fixed-$m$ response according to
\begin{equation}
\mathbf G_{\pm,m}(\omega^*)
=
\mathbf G_{\pm,m}^{*}(\omega).
\label{eq:green_reality_condition}
\end{equation}

More generally, complex conjugation of the mode dependence
$e^{-i\omega t}e^{im\phi}$ relates the paired sectors
$(\omega,m)$ and $(-\omega^*,-m)$. Reflection symmetry gives the direct-sum \mbox{decomposition}
\begin{equation}
\boxed{
\mathbb G(\omega)
=
\mathbf G_+(\omega)
\oplus
\mathbf G_-(\omega).
}
\label{eq:green_parity_direct_sum}
\end{equation}

The direct sum in Equation~\eqref{eq:green_parity_direct_sum} refers to
independent parity sectors; it does not eliminate the off-diagonal
angular-channel components within either block.  The~matrix-valued
response, therefore, retains both the reflection-symmetry organization
of the wormhole and the angular mixing generated by the deformed
geometry.

For visualization, the~real-axis pole singularities may be regulated by
evaluating the response at
\begin{equation}
\omega\rightarrow\omega+i\eta,
\qquad
\eta>0.
\label{eq:green_regulator}
\end{equation}

The regulator $\eta$ is auxiliary and does not represent damping.
For a simple pole,
\begin{equation}
-\operatorname{Im}
\frac{1}{
\omega-\omega_n+i\eta
}
=
\frac{
\eta
}{
(\omega-\omega_n)^2+\eta^2
},
\label{eq:lorentzian_pole}
\end{equation}
so each real normal-mode pole is displayed as a finite Lorentzian
peak.

\begin{figure}[H]
\includegraphics[width=0.92\textwidth]
{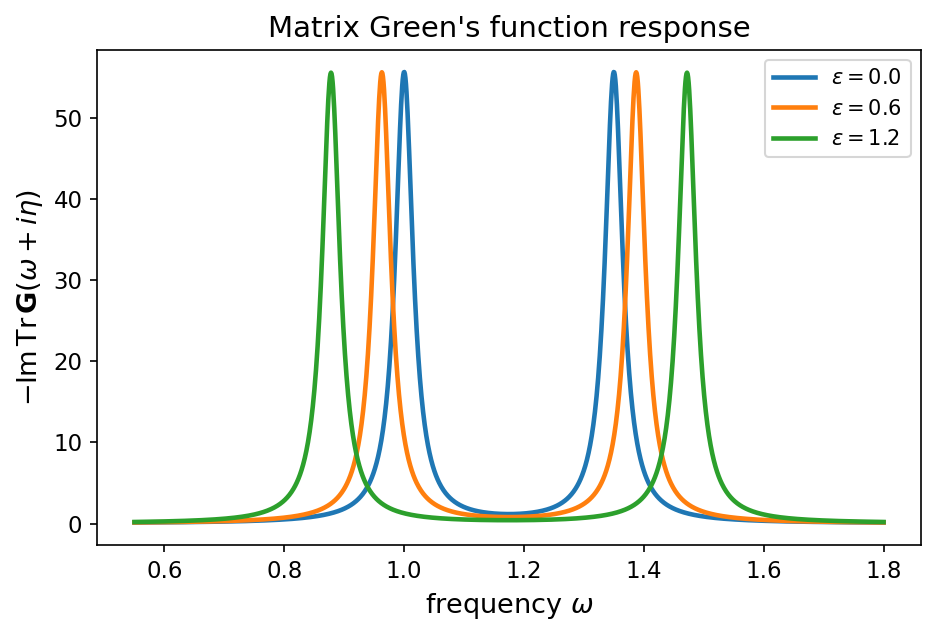}
\caption{Illustrative spectral response
$-\operatorname{Im}\operatorname{Tr}\mathbf G(\omega+i\eta)$
constructed from the reduced two-channel effective model of
Section~\ref{sec:global_spectral}.  This figure is not obtained by
extracting the asymptotic source and response coefficients from the
six-channel numerical bulk solutions.  The~parameter $\eta>0$ is an
auxiliary plotting regulator used to display real normal-mode poles
as finite Lorentzian peaks and does not represent physical damping.
The peaks correspond to poles of the reduced matrix response, whose
locations follow the avoided-crossing branches of
Equation~\eqref{eq:avoided_crossing_roots}.  Increasing the effective
channel coupling separates these poles through the level-repulsion
mechanism of the reduced model.}
\label{fig:green_response}
\end{figure}

The illustrative reduced response model shown in
Figure~\ref{fig:green_response} retains only the angular-channel
structure of the effective two-channel problem discussed in
Section~\ref{sec:global_spectral}; the explicit left/right block
structure of Equation~\eqref{eq:two_boundary_blocks} is suppressed in that
figure.  Its pole locations follow the avoided-crossing branches in
Equation~\eqref{eq:avoided_crossing_roots}.

\begin{figure}[H]

\includegraphics[width=0.95\textwidth]
{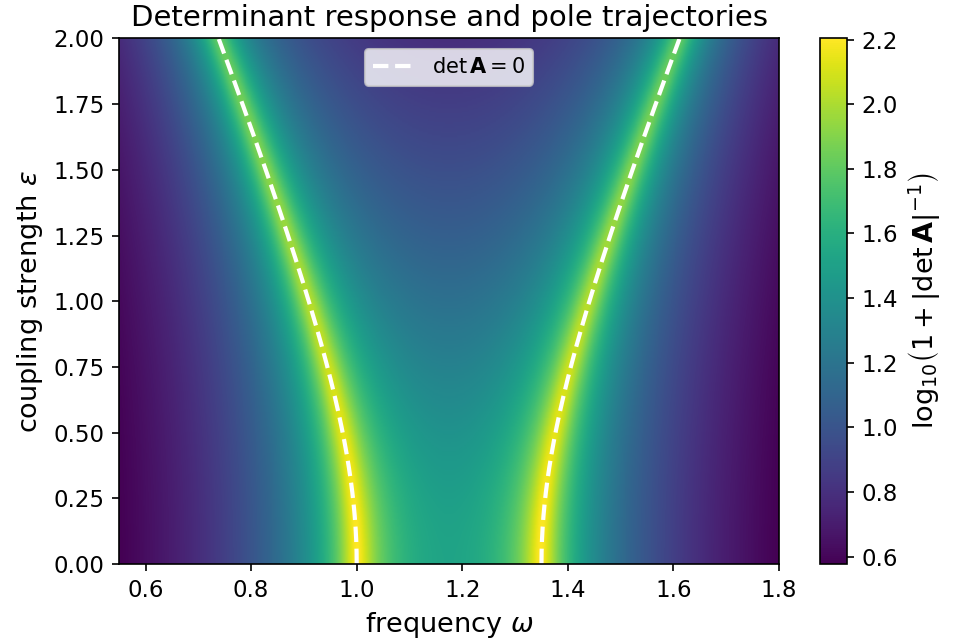}
\caption{Illustrative determinant-response map constructed from the reduced
two-channel effective model.  The~plotted quantity is
$\log_{10}\!\left(1+|\det\mathbf A(\omega+i\eta)|^{-1}\right)$,
so bright ridges occur near zeros of $\det\mathbf A$ and, therefore,
trace the poles of the corresponding reduced matrix response.
The dashed curves show the analytic pole trajectories obtained from
the reduced-model condition $\det\mathbf A(\omega)=0$.  This map is
not obtained from the asymptotic matching matrix of the full
six-channel numerical bulk calculation.  The~parameter $\eta>0$ is
an auxiliary plotting regulator used to render the real-frequency
pole structure and does not represent \mbox{physical damping.}
}
\label{fig:det_response_map}
\end{figure}

The determinant-response diagnostic used in the same illustrative model
is
\begin{equation}
\mathcal D(\omega,\epsilon)
=
\log_{10}
\left[
1+
\left|
\det
\mathbf A_{\rm eff}(\omega+i\eta)
\right|^{-1}
\right],
\label{eq:det_response_diagnostic}
\end{equation}
where $\mathbf A_{\rm eff}$ is the effective two-channel matrix defined
in Equation~\eqref{eq:effective_matching_matrix}.  Large values of
$\mathcal D$ occur near the zeros of the determinant and, therefore,
trace the same pole trajectories as the spectral branches in
Equation~\eqref{eq:avoided_crossing_roots}. The resulting determinant-response map is shown in Figure~\ref{fig:det_response_map}.

A complementary two-boundary observable is provided by the geodesic
approximation for operators of large conformal dimension
~\cite{BalasubramanianRoss2000,LoukoMarolfRoss2000,
AparicioLopez2011,HartmanMaldacena2013}.  For~a heavy bulk field of mass $M_{\rm geo}$, the~worldline path integral is dominated by a spacelike geodesic joining the two boundary insertion points.  The~on-shell worldline action is
\begin{equation}
S_{\rm geo}
=
M_{\rm geo}\,\mathcal L,
\label{eq:geodesic_action}
\end{equation}
where $\mathcal L$ is the proper length of the geodesic.  Using the
large-mass limit of the AdS mass-dimension relation already derived in
Equation~\eqref{eq:AdS_indicial_equation},
\begin{equation}
\Delta\simeq M_{\rm geo}L,
\qquad
M_{\rm geo}L\gg1,
\label{eq:heavy_dimension_limit}
\end{equation}
the bare geodesic contribution becomes
\begin{equation}
\langle O_LO_R\rangle
\sim
\exp
\left(
-\Delta\frac{\mathcal L}{L}
\right).
\label{eq:bare_geodesic_correlator}
\end{equation}

The geodesic length diverges as its endpoints approach the AdS
boundaries.  Introducing a radial cutoff $r_c$, the~universal
asymptotic divergence is removed by
\begin{equation}
\mathcal L_{\rm reg}
=
\lim_{r_c\rightarrow\infty}
\left[
\mathcal L(r_c)
-
2L
\log
\left(
\frac{2r_c}{L}
\right)
\right].
\label{eq:renormalized_geodesic_length}
\end{equation}

The logarithmic term follows directly from the large-$r$ behavior
already established in Equations~\eqref{eq:grr_ads_asymptotic} and
\eqref{eq:ell_of_r}; no additional asymptotic geometry is required
~\cite{FestucciaLiu2006}.

The renormalized two-boundary correlator is, therefore,
\begin{equation}
\boxed{
\langle O_LO_R\rangle_{\Delta\gg1}
\sim
\exp
\left(
-\Delta
\frac{
\mathcal L_{\rm reg}
}{L}
\right).
}
\label{eq:geodesic_correlator}
\end{equation}

A spacelike geodesic connecting the two asymptotic boundaries probes
the global wormhole geometry, so
$\mathcal L_{\rm reg}$ measures the effective bulk separation between
the two conformal regions.  Variations of the geometry modify
$\mathcal L_{\rm reg}$ and are exponentially amplified in
Equation~\eqref{eq:geodesic_correlator} when $\Delta\gg1$.

The response operator
Equation~\eqref{eq:global_two_boundary_response} and the correlator
Equation~\eqref{eq:geodesic_correlator} probe complementary aspects of the
same two-ended spacetime.  The~former provides the formal
matrix-valued framework relating boundary sources and responses and
encodes the collective spectral organization and angular-channel
mixing of the scalar field, whereas the latter probes the global
geometric connectivity of the wormhole.  The~local quantum observable
considered next is formulated in terms of the same coupled normal-mode
structure.


We finally examine the formal construction of the local observable
$\langle\Phi^2(x)\rangle_{\rm ren}$ and the role of angular-channel
mixing in its mode-sum representation.  The~analysis below derives
the local Hadamard singular structure and its geometry-dependent
short-distance subtraction, and~then illustrates the finite
inter-channel contribution within the coupled-mode representation.
It does not constitute a complete numerical evaluation of
$\langle\Phi^2(x)\rangle_{\rm ren}$, which would additionally require
specification of the quantum state, evaluation of the complete
spectral mode sum, inclusion of all geometry-dependent subtraction
terms required at coincidence, and~demonstration of convergence of
the regularized~result.

Let $\sigma$ label the positive-frequency coupled modes entering the
formal mode expansion.  Using the harmonic basis of
Equation~\eqref{eq:harmonic_expansion_ads}, write
\begin{equation}
u_\sigma(x)
=
\mathcal N_\sigma
e^{-i\omega_\sigma t}
e^{im\phi}
\sum_L
C_{\sigma L}
R_{\sigma L}(\ell)
Y_{Lm}(\theta),
\label{eq:coupled_quantum_mode}
\end{equation}
where $C_{\sigma L}$ are the angular-channel components of the
normalized eigenvector obtained from the global coupled spectral
problem.  The~positive-frequency Wightman function is
\begin{equation}
G^+(x,x')
=
\sum_\sigma
u_\sigma(x)
u_\sigma^*(x'),
\label{eq:coupled_Wightman}
\end{equation}
and, therefore, inherits the matrix structure in angular-channel space,
\begin{equation}
G^+(x,x')
=
\sum_{L,L'}
Y_{Lm}(\theta,\phi)\,
G^+_{LL'}(t,\ell;t',\ell')\,
Y^*_{L'm}(\theta',\phi').
\label{eq:Wightman_matrix}
\end{equation}

The terms with $L\neq L'$ contain the local interference between
angular channels generated by the non-separable~geometry.

For a Hadamard state in four spacetime dimensions, the~singular
two-point function has the local form
\begin{equation}
G_{\rm sing}(x,x')
=
\frac{1}{8\pi^2}
\left[
\frac{U(x,x')}{\sigma(x,x')}
+
V(x,x')
\ln
\left(
\mu_{\rm ren}^2\sigma(x,x')
\right)
\right],
\label{eq:Hadamard_singular}
\end{equation}
with $U(x,x')=1+\mathcal O(\sigma)$ in the coincidence limit
~\cite{DeWittBrehme1960,Christensen1976,
DecaniniFolacci2008,AndersonHiscockSamuel1995}.
The renormalized vacuum polarization is
\begin{equation}
\boxed{
\langle\Phi^2(x)\rangle_{\rm ren}
=
\lim_{x'\rightarrow x}
\left[
G^+(x,x')
-
G_{\rm sing}(x,x')
\right].
}
\label{eq:vacuum_polarization_definition}
\end{equation}

For the stationary mode basis used here, we regulate the coincidence
limit by an imaginary displacement along the Killing-time direction at
fixed spatial coordinates,
\begin{equation}
x^\mu
=
(t,r,\theta,\phi),
\qquad
x'^\mu
=
(t+i\delta,r,\theta,\phi),
\qquad
\delta>0.
\label{eq:fixed_phi_splitting}
\end{equation}

The mode phase then contributes
\begin{equation}
e^{-i\omega_\sigma(t-t')}
=
e^{-\omega_\sigma\delta},
\label{eq:temporal_regulator}
\end{equation}
so this splitting is the coordinate-space realization of the
exponential regulator appearing in the spectral~sum.

The required local metric factor is already contained in
Equation~\eqref{eq:gtt_exact}.  Define
\begin{equation}
\chi^2(r,\theta)
\equiv
-g_{tt}
=
N^2(r,\theta)
-
r^2\Omega^2(r)\sin^2\theta.
\label{eq:chi_definition}
\end{equation}

The ergoregion-free condition imposed throughout the calculation
guarantees
\begin{equation}
\chi^2(r,\theta)>0.
\label{eq:chi_positive}
\end{equation}

For the displacement in Equation~\eqref{eq:fixed_phi_splitting}, the~local
expansion of Synge's world function gives
\begin{equation}
2\sigma(x,x')
=
g_{tt}(-i\delta)^2
+
\mathcal O(\delta^4)
=
\chi^2(r,\theta)\delta^2
+
\mathcal O(\delta^4),
\label{eq:sigma_temporal}
\end{equation}
or
\begin{equation}
\sigma(x,x')
=
\frac{1}{2}
\chi^2(r,\theta)\delta^2
+
\mathcal O(\delta^4).
\label{eq:sigma_temporal_half}
\end{equation}

Thus, the local proper temporal separation is
\begin{equation}
\delta_{\rm proper}
=
\chi(r,\theta)\delta.
\label{eq:proper_temporal_separation}
\end{equation}

Substitution of Equation~\eqref{eq:sigma_temporal_half} into
Equation~\eqref{eq:Hadamard_singular} gives the leading singular term
\begin{equation}
\boxed{
G_{\rm sing}(\delta;r,\theta)
=
\frac{1}{
4\pi^2\chi^2(r,\theta)\delta^2
}
+
\mathcal O(\ln\delta).
}
\label{eq:geometry_Hadamard_subtraction}
\end{equation}

Using the frame-dragging function already specified in
Equation~\eqref{eq:metric_functions},
\begin{equation}
\boxed{
G_{\rm sing}(\delta;r,\theta)
=
\frac{1}{
4\pi^2
\left[
N^2(r,\theta)
-
\dfrac{4a^2}{r^4}\sin^2\theta
\right]
\delta^2
}
+
\mathcal O(\ln\delta).
}
\label{eq:explicit_Hadamard_subtraction}
\end{equation}

The subtraction, therefore, retains the local redshift and
frame-dragging information of the rotating geometry.  In~terms of the
proper separation in Equation~\eqref{eq:proper_temporal_separation}, the~leading ultraviolet singularity retains its universal~form.

At the throat, Equations~\eqref{eq:lapse_throat} and
\eqref{eq:metric_functions} give
\begin{equation}
\chi_0^2(\theta)
=
1
-
\frac{4a^2}{r_0^4}
\sin^2\theta,
\label{eq:chi_throat}
\end{equation}
and hence
\begin{equation}
\boxed{
G_{\rm sing}^{\rm throat}(\delta,\theta)
=
\frac{1}{
4\pi^2
\left[
1-\dfrac{4a^2}{r_0^4}\sin^2\theta
\right]
\delta^2
}
+
\mathcal O(\ln\delta).
}
\label{eq:throat_Hadamard_subtraction}
\end{equation}

The same factor controlling the ergoregion condition, therefore, controls
the relation between Killing-time splitting and the local proper
separation.  The~bound in
Equation~\eqref{eq:ergoregion_global_bound} ensures that the fixed-$\phi$
temporal regulator remains locally well defined throughout the
parameter domain used in the spectral calculation.  The~original
co-rotating comparison is unnecessary here: the regulator
Equation~\eqref{eq:temporal_regulator} fixes the appropriate splitting
prescription~directly.

At coincidence, Equation~\eqref{eq:coupled_Wightman} gives the regulated
mode sum
\begin{equation}
G_\delta^+(x,x)
=
\sum_\sigma
|\mathcal N_\sigma|^2
e^{-\omega_\sigma\delta}
\left|
\sum_L
C_{\sigma L}
R_{\sigma L}(\ell)
Y_{Lm}(\theta)
\right|^2.
\label{eq:regulated_Wightman_coincident}
\end{equation}

Expanding the channel sum,
\begin{equation}
\begin{aligned}
G_\delta^+(x,x)
={}&
\sum_{\sigma,L}
|\mathcal N_\sigma|^2
e^{-\omega_\sigma\delta}
|C_{\sigma L}|^2
|R_{\sigma L}|^2
|Y_{Lm}|^2
\\
&+
2\,\mathrm{Re}
\sum_{\sigma}
\sum_{L<L'}
|\mathcal N_\sigma|^2
e^{-\omega_\sigma\delta}
C_{\sigma L}
C_{\sigma L'}^*
R_{\sigma L}
R_{\sigma L'}^*
Y_{Lm}
Y_{L'm}^* .
\end{aligned}
\label{eq:Wightman_diag_offdiag}
\end{equation}

The first line contains the diagonal channel contributions; the second
contains the interference terms produced by the coupled
eigenvectors.

Combining Equations~\eqref{eq:vacuum_polarization_definition},
\eqref{eq:geometry_Hadamard_subtraction}, and~\eqref{eq:Wightman_diag_offdiag} gives
\begin{equation}
\boxed{
\begin{aligned}
\langle\Phi^2(x)\rangle_{\rm ren}
=
\lim_{\delta\rightarrow0}
\Bigg\{&
\sum_{\sigma,L}
|\mathcal N_\sigma|^2
e^{-\omega_\sigma\delta}
|C_{\sigma L}|^2
|R_{\sigma L}|^2
|Y_{Lm}|^2
\\
&+
2\,\mathrm{Re}
\sum_{\sigma}
\sum_{L<L'}
|\mathcal N_\sigma|^2
e^{-\omega_\sigma\delta}
C_{\sigma L}
C_{\sigma L'}^*
R_{\sigma L}
R_{\sigma L'}^*
Y_{Lm}
Y_{L'm}^*
\\
&-
G_{\rm sing}(\delta;r,\theta)
\Bigg\}.
\end{aligned}
}
\label{eq:vacuum_polarization_coupled}
\end{equation}

Equation~\eqref{eq:vacuum_polarization_coupled} displays the formal
coupled-channel structure of the Hadamard-subtracted coincidence
limit.  A~complete evaluation of
$\langle\Phi^2(x)\rangle_{\rm ren}$ would require the full
state-dependent spectral sum together with all subtraction terms
needed to obtain a finite, converged coincidence limit.  In~the
finite-channel illustration considered below, the~off-diagonal terms
instead provide a representation-dependent measure of interference
among the retained spherical-harmonic components.  They vanish when
the spectral problem is diagonal in the chosen $L$ basis and survive
when the non-separable geometry generates coupled-channel amplitudes.
The throat parity conditions need not be re-derived here.  From~Equations~\eqref{eq:even_throat_bc} and \eqref{eq:odd_throat_bc}, odd modes
vanish at $\ell=0$, whereas even modes may remain finite there.
Consequently, the~local interference structure of
Equation~\eqref{eq:vacuum_polarization_coupled} is parity dependent near the
throat.

\begin{figure}[H]

\includegraphics[width=0.90\textwidth]
{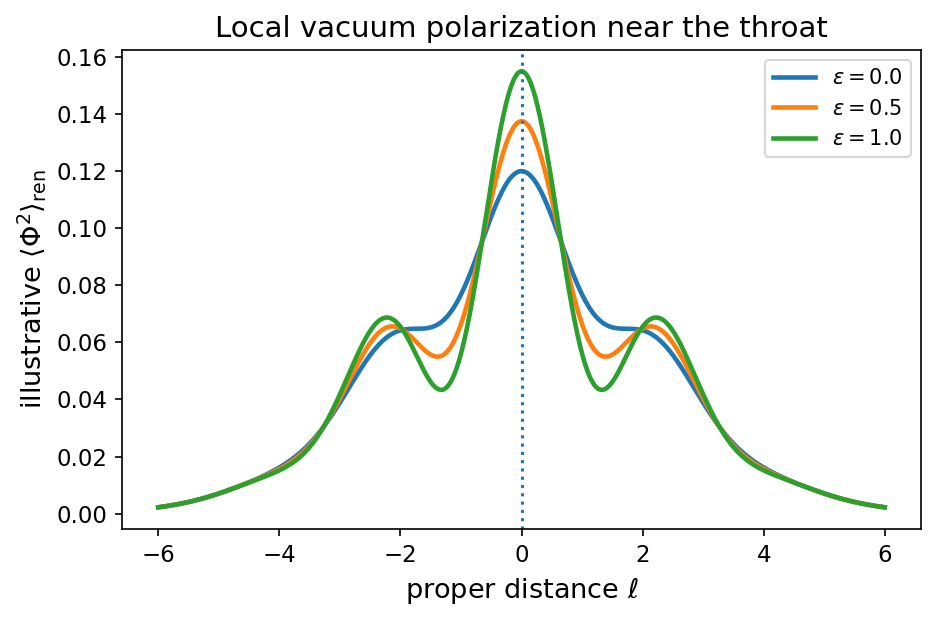}
\caption{Illustrative finite-channel contribution to the local mode-sum
structure near the wormhole throat for several values of the
angular-channel coupling strength $\epsilon$.  The~plotted curves
retain the finite coupled-channel contribution after subtraction of
the leading universal Hadamard short-distance singularity.  Diagonal
terms provide the baseline profile, while off-diagonal terms describe
interference among components in the chosen spherical-harmonic basis.
The figure is a finite-channel illustration and is not a complete
calculation of $\langle\Phi^2\rangle_{\rm ren}$: no complete
state-dependent spectral sum, full geometry-dependent Hadamard
subtraction, or~convergence analysis of the regularized coincidence
limit is performed here. The~vertical dotted line marks the
wormhole throat.}
\label{fig:vacuum_polarization_profile}
\end{figure}

Figure~\ref{fig:vacuum_polarization_profile} illustrates only the finite
interference mechanism contained in
Equation~\eqref{eq:vacuum_polarization_coupled}; it should not be interpreted
as a full numerical calculation of
$\langle\Phi^2\rangle_{\rm ren}$.
The result established here is instead structural: the local geometry
determines the short-distance Hadamard structure entering
Equation~\eqref{eq:explicit_Hadamard_subtraction}, while the non-separable
spectral problem generates diagonal and off-diagonal contributions in
the chosen spherical-harmonic representation.  The~finite-channel
calculation displayed here isolates the corresponding interference
structure but does not evaluate the complete state-dependent,
Hadamard-renormalized coincidence~limit.

\section{Discussion and~Conclusions}
\label{sec:discussion_conclusions}

We have developed a geometry-to-spectrum construction for a probe
scalar field in a rotating asymptotically AdS-Teo wormhole.  The~central result is not the observation that the scalar equation becomes
non-separable.  Rather, the~controlled loss of separability provides
the starting point for a global coupled-channel problem in which the
spacetime geometry determines the matrix differential operator, the~AdS boundary conditions determine its spectral realization, and~the
same coupled spectral framework organizes the formal two-boundary
response and the channel structure entering local quantum mode sums.
This formulation is complementary to existing Ellis--Bronnikov
studies: explicit asymptotically AdS wormhole solutions have been
constructed from specified matter models
~\cite{BlazquezSalcedo2021AdS}, while rotating configurations have been
analyzed using slow-rotation, radial and quadrupolar perturbations, and~quasinormal-mode calculations
~\cite{Azad2024SlowRotation,Azad2024SecondOrder,Azad2025Quadrupole,Khoo2024QNM}.
Here, by~contrast, the~emphasis is on a prescribed rotating AdS-Teo
background and on the geometry-derived non-separable matrix operator
governing the probe-scalar spectral problem.  The~first question posed
in the Introduction concerned the geometry: whether a rotating
Teo-type throat can be incorporated into a consistent asymptotically
AdS background while retaining a controlled source of angular
non-separability and an admissible regime without an ergoregion.  The~construction in Section~\ref{sec:geometry_scalar} answers this
explicitly.  The~radial factor in Equation~\eqref{eq:A_definition} changes
the asymptotically flat Teo radial sector so that the exact coefficient
Equation~\eqref{eq:grr_exact} approaches the required AdS behavior
Equation~\eqref{eq:grr_ads_asymptotic}, while the throat conditions remain
regular.  The~proper-distance coordinate Equation~\eqref{eq:r_of_ell}
covers both exterior regions continuously with the throat at
$\ell=0$.

The angular dependence is likewise an explicit part of the geometry
rather than formal notation.  The~deformation
Equation~\eqref{eq:lapse_deformation} vanishes at the throat and
asymptotically but is nonzero in the intermediate region.  Its
quadrupolar structure produces the selection rule
Equation~\eqref{eq:selection_rule_ads} and hence directly generates the
off-diagonal channel couplings.  The~non-separable scalar problem is,
therefore, tied to a definite metric deformation whose geometric origin
can be followed through every subsequent~projection.

Rotation introduces an additional physical restriction.  Horizonlessness
alone does not exclude an ergoregion, so the spectral calculation was
not performed over unrestricted values of $a$ and $\epsilon$.  The~analysis of $g_{tt}$ leads to the sufficient global condition
Equation~\eqref{eq:ergoregion_global_bound}.  All normal-mode calculations
reported here lie within this ergoregion-free domain.  The~absence of
resolved complex frequencies within numerical precision and the
conservative interpretation of the computed spectrum should, therefore,
be understood as properties of the parameter regime, discretizations,
and boundary-value problem studied in this paper, not as an exact
analytical statement about all rotating horizonless wormholes or all
quadratic spectral~pencils.

The second question concerned what replaces the ordinary separated
radial eigenvalue problem once the angular dependence of the geometry
couples different harmonics.  Projection of the Klein--Gordon equation
produces the geometry-derived matrices
$\mathbf P$, $\mathbf W$, $\mathbf C$, and~$\mathbf H$ defined in
Equations~\eqref{eq:P_matrix_ads}--\eqref{eq:H_matrix_ads}.  The~resulting
radial dynamics are, therefore, a matrix Sturm--Liouville differential
system rather than a collection of independent equations.  In~the
explicit $m=1$, $L\in\{1,3\}$ realization, the~overlap matrices and all
off-diagonal coefficients were calculated directly from the
quadrupolar metric deformation; no additional channel-coupling
potential was~introduced.

\textls[-25]{The mathematical spectral problem also contains an important
distinction.  The~frequency-independent} differential operator
Equation~\eqref{eq:A_operator}, together with its domain,
Equation~\eqref{eq:operator_domain}, is distinct from the
frequency-dependent quadratic operator pencil
Equation~\eqref{eq:quadratic_operator_pencil}.  Throat regularity,
reflection parity, and~the normalizable AdS falloff determine the
admissible boundary conditions.  Propagating a complete basis of
throat-regular solutions to the asymptotic region then produces the
matching matrices in Equation~\eqref{eq:matching_matrices_AB}.  The~physical
normal modes occur when the source matrix loses rank, giving the
determinant condition Equation~\eqref{eq:global_determinant_condition}.
Thus, the normal-mode spectrum is a property of the complete
geometry-derived boundary-value problem and not of an auxiliary
two-state~model.

This construction also clarifies the role of the two-channel analysis.
The explicit two-channel determinant
Equation~\eqref{eq:two_channel_quantization} provides the smallest
nontrivial realization of the geometry-derived problem and exposes
analytically how coupling produces frequency shifts and level
repulsion.  The~avoided-crossing formula
Equation~\eqref{eq:avoided_crossing_roots} is, therefore, useful for
interpretation, but~the numerical conclusions of the paper do not rest
on \mbox{that~approximation.}

 The third question was whether the finite harmonic truncation actually
converges and whether the spectral structures identified in the
minimal model survive when additional channels are retained.  The~numerical studies in Section~\ref{sec:numerical_spectrum} provide a
direct answer.  The~radial-resolution test at fixed finite domain shows
systematic convergence as $N_r$ is increased, while the combined
finite-domain and radial-resolution test shows systematic numerical
convergence of the low-lying spectrum as $\ell_{\max}$ and $N_r$ are
enlarged together; the latter is not an independent cutoff-convergence
test.  The~two-, four-, and~six-channel calculations then test
convergence under enlargement of the angular basis.  The~successive
relative changes defined in Equation~\eqref{eq:channel_relative_changes}
decrease systematically from the two-to-four-channel comparison to the
four-to-six-channel comparison.  For~the representative low-lying
spectrum in Table~\ref{tab:channel_convergence}, the~four- and
six-channel frequencies agree to substantially higher accuracy than
the two- and four-channel results, with~the remaining corrections
ranging from approximately $10^{-11}$ to $10^{-5}$ over the displayed
modes.  The~low-lying normal-mode frequencies are, therefore,
numerically converged under the tested radial and angular refinements
and show systematic numerical convergence under simultaneous
enlargement of the finite radial domain and radial~basis.

The angularly converged calculation also sharpens the physical
interpretation of the level-repulsion structure.  At~the production
resolution $(\ell_{\max},N_r)=(25,45)$, the~six-channel rotation
continuation gives a finite interior minimum separation at
$a_\star\simeq0.3943$, with~$\Delta\omega_{\min}\simeq1.689\times10^{-3}$.  The~dedicated
resolution audit shows that this feature converges toward
$a_\star\simeq0.3951$ and
$\Delta\omega_{\min}\simeq1.673\times10^{-3}$ at the highest
tested resolution $(\ell_{\max},N_r)=(30,60)$, well before the
ergoregion-free boundary is reached.  The~corresponding channel-weight
evolution shows that the participating eigenvectors are already
multichannel objects on both sides of the minimum and that their
angular composition is redistributed continuously through the
minimum-gap region.  The~numerical result is, therefore, most accurately
described as a collective multichannel spectral reorganization with
finite level repulsion.  The~reduced two-level model captures the
qualitative local mechanism by which coupling prevents an exact
crossing, whereas the six-channel calculation determines the actual
branch trajectories and multichannel eigenvector~composition.

This convergence result is important conceptually. Once the geometry
contains nonzero off-diagonal angular projections, a~single harmonic
cannot be assumed to define an autonomous physical mode.  The~fundamental object is the coupled operator, and~any finite collection
of harmonics is a Galerkin approximation whose validity must be tested.
The two-, four-, and~six-channel comparison provides that test
for the low-lying spectrum considered~here.

The fourth question concerned what information the same coupled
spectral framework carries beyond the eigenfrequencies themselves.
The two asymptotic AdS regions require independent left and right
boundary data, leading formally to the block response operator
Equation~\eqref{eq:two_boundary_blocks}, rather than a single-boundary
scalar Green function.  Its diagonal boundary blocks represent
same-boundary response, while its off-diagonal blocks represent
cross-boundary transmission through the wormhole.  Within~each block,
the angular-channel indices retain the non-separable structure
generated by the bulk geometry.  This construction defines the formal
two-boundary response structure; the full block response is not
numerically reconstructed from the six-channel bulk solutions in this~work.

In the parity basis the formal response takes the form
Equation~\eqref{eq:parity_green_function}, with~poles determined by the same
matching condition that selects the normal modes and residues encoding
the channel structure of the corresponding collective excitations.
The determinant spectrum and matrix-valued response are, therefore,
formally complementary representations of the same global spectral
problem~\cite{SonStarinets2002,KaminskiOperatorMixing2010}.  In~this
work, however, the~full response matrix is not numerically
reconstructed from the asymptotic coefficients of the six-channel
bulk solutions.  The~response plots instead illustrate this pole
structure using the reduced two-channel effective model, whereas the
geometry-derived six-channel calculation establishes the normal-mode
spectral flow and resolves no complex frequencies within numerical
precision over the tested ergoregion-free~regime.

The local quantum calculation provides a second manifestation of the
same channel structure.  The~Wightman function
Equation~\eqref{eq:Wightman_matrix} contains diagonal and off-diagonal
angular contributions.  Hadamard renormalization requires the local
short-distance subtraction to be derived from the actual rotating
metric.  For~the fixed-$\phi$ Killing-time splitting used here,
Equations~\eqref{eq:sigma_temporal_half} and
\eqref{eq:explicit_Hadamard_subtraction} show that the regulator is
controlled by
$\chi^2=-g_{tt}$ rather than by the lapse alone.  The~same metric
quantity that enters the ergoregion analysis, therefore, determines the
conversion between coordinate-time separation and local proper
separation in the ultraviolet~subtraction.

After removal of the leading universal short-distance singularity,
the formal coupled-channel expression
Equation~\eqref{eq:vacuum_polarization_coupled} contains diagonal and
off-diagonal contributions constructed from the retained angular-mode
amplitudes.  The~distinction between the separable and non-separable
regimes is explicit in the geometry-derived operators: for
$\epsilon=0$, the~deformation in Equation~\eqref{eq:lapse_deformation}
vanishes and the associated off-diagonal channel couplings disappear,
whereas for $\epsilon\neq0$ the overlap structure of
Equation~\eqref{eq:P2_overlap_general} generates finite inter-channel
mixing.  The~resulting off-diagonal terms provide a
representation-dependent diagnostic of interference among components
in the chosen spherical-harmonic basis rather than, by~themselves, a~basis-independent local quantum observable.  The~calculation
establishes the coupled-channel structure and the local short-distance
subtraction entering the formal Hadamard construction, while the
profile in Figure~\ref{fig:vacuum_polarization_profile} is intentionally
a finite-channel illustration and not a complete numerical evaluation
of $\langle\Phi^2\rangle_{\rm ren}$.

The cross-boundary geodesic correlator
Equation~\eqref{eq:geodesic_correlator} provides a complementary geometric
probe.  Whereas the matrix Green function measures the spectral
response of coupled bulk modes, the~renormalized geodesic length
measures the global spacelike connection between the two asymptotic
regions in the large-conformal-dimension approximation.  We, therefore,
do not identify the geodesic construction with the full spectral
response; the two observables probe different aspects of the same
two-ended AdS~geometry.

Symmetry remains the organizing principle throughout this construction
even though complete separability is absent.  Stationarity and
axisymmetry preserve the Killing labels $\omega$ and $m$; throat
reflection symmetry decomposes the global problem into independent
parity sectors; the quadrupolar deformation preserves equatorial
reflection symmetry and restricts the coupled harmonic hierarchy
through Equation~\eqref{eq:selection_rule_ads}; and the operator and
boundary conditions determine the conservative spectral realization
within the admissible parameter regime.  Controlled non-separability,
therefore, does not amount to an absence of symmetry.  It replaces the
stronger symmetry responsible for complete separation by a residual
symmetry structure that organizes a matrix-valued spectral~problem.

Several limitations delimit the claims of this work.  The~rotating AdS-Teo spacetime is treated as a phenomenological fixed
background supported by the effective source
\mbox{Equation~\eqref{eq:effective_stress_tensor};} we do not derive a microscopic
Einstein-matter model for the wormhole.  The~scalar is a probe field,
so the absence of growing modes in the calculations reported here
does not establish gravitational stability of the supporting geometry
or its matter sector.  The~ergoregion condition used here is a
sufficient restriction defining the domain of the calculations, not a
complete stability theorem for every rotating configuration.  The~convergence study establishes numerical convergence of the retained
low-lying normal-mode frequencies over the channel spaces examined,
but does not replace a systematic infinite-channel analysis in
strongly mixed regimes.  Finally, the~full two-boundary response
matrix is not numerically reconstructed from the six-channel bulk
solutions, and~the finite-channel profile in
Figure~\ref{fig:vacuum_polarization_profile} is not a complete numerical
evaluation of the Hadamard-renormalized
$\langle\Phi^2\rangle_{\rm ren}$.  These distinctions delimit the
claims established here and identify natural extensions of \mbox{the
framework.}

The four questions posed in the Introduction, therefore, have concrete
answers.  A~regular rotating Teo throat can be joined to a radial
sector with the required AdS asymptotics while supporting an explicit
localized angular deformation and an ergoregion-free working domain.
The resulting non-separable scalar dynamics are governed by
geometry-derived coupled operators and a quadratic spectral pencil,
with global normal modes selected by the loss of invertibility of the
AdS matching matrix.  The~finite-channel realization is numerically
converged for the low-lying modes examined here, and~the six-channel
rotation continuation exhibits finite level repulsion accompanied by
a collective redistribution of the multichannel eigenvectors.
Finally, the~same coupled spectral framework provides the formal
construction of a two-boundary matrix response and organizes the
coupled-channel structure entering local quantum mode sums.  The~response and local quantum profiles presented here are reduced-model
and finite-channel illustrations, respectively, rather than complete
numerical evaluations of the full six-channel boundary response or
of $\langle\Phi^2\rangle_{\rm ren}$.

The principal outcome is, therefore, the following unified chain: the
geometry fixes the coupled operator, the~operator and global boundary
conditions fix the spectrum, and~the same coupled spectral framework
organizes both the formal asymptotic response and the channel structure
entering local quantum mode sums.  In~this sense, the~normal modes,
level-repulsion structure, boundary-response poles, and~inter-channel
terms arise from the same geometry-induced coupled-channel framework,
although the response and local quantum quantities are not evaluated
here at the same numerical level of completeness as the six-channel
normal-mode~spectrum.

The rotating AdS-Teo wormhole consequently provides a concrete
example in which the appropriate replacement for complete mode
separability is not an approximate single-channel equation, but~a
symmetry-organized matrix spectral theory.  More generally, the~results
suggest that in non-separable rotating geometries the coupled operator
itself should be regarded as the fundamental dynamical object, with~single-channel descriptions justified only as controlled projections
of that underlying~system.

\vspace{+6pt}
\section*{Author Contributions}
 Conceptualization, Ramesh Radhakrishnan and Gerald Cleaver; Methodology, Ramesh Radhakrishnan and Gerald Cleaver; Software, Ramesh Radhakrishnan; Validation, Ramesh Radhakrishnan, William Julius and Gerald Cleaver; Formal analysis, Ramesh Radhakrishnan, William Julius and Gerald Cleaver; Investigation, Ramesh Radhakrishnan, William Julius and Gerald Cleaver; Resources, Ramesh Radhakrishnan and Gerald Cleaver; Data curation, Ramesh Radhakrishnan, William Julius and Gerald Cleaver; Writing – original draft, Ramesh Radhakrishnan; Writing – review and editing, Ramesh Radhakrishnan, William Julius and Gerald Cleaver; Visualization, Ramesh Radhakrishnan and Gerald Cleaver; Supervision, Gerald Cleaver; Project administration, Gerald Cleaver. All authors have read and agreed to the published version of the manuscript.

\section*{Funding}
This research received no external funding.

\section*{Data Availability}
No new data were created or analyzed in this study. Data sharing is not applicable to this article.

\section*{Conflicts of Interest}
The authors declare no conflict of interest.



\end{document}